\documentclass[aps,11pt,eqsecnum, preprint,nofootinbib,superscriptaddress]{revtex4}
\usepackage{amssymb,amsmath,amsthm,graphicx,amscd}
\usepackage{enumerate,xcolor,changepage,ulem,bm}
\usepackage[hidelinks]{hyperref}

\begin{document} 
\title{No intrinsic decoherence of \\ inflationary cosmological perturbations} 
\author{Jen-Tsung Hsiang}
\email{cosmology@gmail.com}
\affiliation{Center for High Energy and High Field Physics, National Central University, Taoyuan 320317, Taiwan, ROC}
\author{Bei-Lok Hu}
\email{blhu@umd.edu}
\affiliation{Maryland Center for Fundamental Physics and Joint Quantum Institute,  University of Maryland, College Park, Maryland 20742, USA}

\begin{abstract}
After a brief summary of the four main veins since the 80s in the treatment of decoherence and quantum to classical transition in cosmology we  focus on one of these veins in the study of quantum decoherence of cosmological perturbations in inflationary universe,  the  case when it does not rely on any environment.  This is what `intrinsic' in the title refers to -- a closed quantum system, consisting of a quantum field which drives inflation.  The question is, does its quantum perturbations, that which interact with the density contrast giving rise to structures in the universe, decohere with an inflationary expansion of the universe. A dominant view which had propagated for a quarter of a century asserts  yes,  based on the belief that the large squeezing of a quantum state after a duration of inflation renders  the system effectively classical.   This paper  debunks this view by identifying the technical fault-lines in its derivations  and revealing the pitfalls in its arguments which drew earlier authors to this wrong conclusion. We use a few simple quantum mechanical models to expound where the fallacy originated:  The highly squeezed ellipse quadrature in phase space cannot be simplified to a line, and the Wigner function cannot be replaced by a delta function.  These measures amount to taking only the leading order in the relevant parameters in seeking the semiclassical limit and ignoring the subdominant contributions where quantum features reside. Doing so violates the bounds of the Wigner function, and its wave functions possess negative eigenvalues. Moreover, the Robertson-Schr\"odinger uncertainty relation for a pure state is violated.  For  inflationary cosmological perturbations,  in addition to these features,  entanglement exists between the created pairs. This uniquely quantum feature cannot be easily argued away. Indeed it could be our best hope to retroduce the quantum nature of cosmological perturbations and the trace of an inflaton field.  All this point to the invariant fact that a closed quantum system, even when highly squeezed, evolves unitarily without loss of coherence; quantum cosmological perturbations do not by itself decohere.

 
--{\it Invited paper for Universe: Special Issue on ``Quantum Aspects of the Universe". }
\end{abstract}

\maketitle

\hypersetup{linktoc=all}

\baselineskip=18pt

\allowdisplaybreaks

\newpage

\section{Introduction}

Long before the 1996 popular paper \cite{PS96} of Polarski and Starobinsky (PS) on cosmological decoherence which commanded the attention of  the cosmology community since then,  there had already been intense activities on the issue of quantum  decoherence:  two major paradigms of consistent/decoherent histories \cite{Griffiths,Griffiths1,Omnes,Omnes1,Omnes2,GelHar89,Hartle90,GelHar93}  and the environment-induced decoherence \cite{Zurek,JoosZeh,JoosZehBook,Schloss,Schloss1} programs had been formulated and the emergence of classical spacetime in quantum cosmology \cite{HallQC} investigated. The theoretical foundation for  quantum decoherence, laid down from the early 80s to the early 90s,  is essential for understanding the corresponding issues of cosmological perturbations in inflationary cosmology. Even the first step in the analysis, e.g., whether to take a closed quantum system, such as pursued by Hartle \cite{HartleClosed1,HartleClosed2} or an open quantum system \cite{qos,qos1,qos2} viewpoint, makes a big difference. For a closed system we can just do quantum mechanics the usual way, as had been done earlier.  What   is different as presented in this paper is, we shall identify the fault-lines and pitfalls which tricked some earlier authors to jump to wrong conclusions.  From  an open system perspective,  it is a welcoming sign that in recent years  nonequilibrium quantum field theory \cite{NEqFT,NEqFT1,NEqFT2,NEqFT3} developed in the 80s-90s are increasingly recognized to be essential for a rigorous treatments of cosmological decoherence issues.  Just so that beginning researchers would not overlook the complexity of this issue arduously worked out by authors of that earlier period,  and take an easy-does-it attitude 
to this issue,  we give an overview of the four stages and the four veins of quantum decoherence research in gravitation and cosmology in the 80s and 90s, up to PS. Because of its pedagogical emphasis we place it in a separate section, Sec.~\ref{S:erjbd},  with a remark that readers familiar with this historical development can skip over.

The main goal of this paper is to focus on one of these veins in the study of quantum decoherence of cosmological perturbations in inflationary universe,  the  case when one does not rely on an environment.  This is what `intrinsic' in the title refers to -- a closed system. Namely, a quantum field which drives inflation by itself, the inflaton, no environment.  In fact, a free field\footnote{Here, the center of attention is the quantum perturbations of the inflaton field,  that is, $\delta\hat{\varphi}(\bm{x},t)$ (see next subsection); the mean field $\bar\phi(t)$ 
is governed by a potential $V(\bar\phi)$ which engineers the inflationary dynamics, such as `slow-roll', etc.}  The cases of an interacting field, where one divides the high frequency modes from the low frequency modes, and examine how the former sector decoheres the latter,  had been studied in details before \cite{CalHu95,LomMaz96,LomNac05} and  gathering increasing momentum in recent papers (see, e.g., \cite{Nelson,HolMcD17,Brahma} and references therein). Despite its simplicity the issue of decoherence for a free field is perhaps conceptually  more challenging, not unlike defining  the `intrinsic' entropy of a quantum field (see, e.g.,~\cite{IntEnt} and earlier references cited there in). This is because,  if one adheres to the basic principles, a closed quantum system should evolve unitarily -- there is no loss of quantum coherence.  What makes this an issue is because there are papers by respectable authors which claim there is decoherence in this closed quantum system at late times, due to the inflationary expansion.  Because PS offered  an easy, simple, even faulty philosophically sophisticated  explanation of decoherence   (`decoherence without decoherence'),  it has attracted a large numbers of followers.  
We want to show in this paper that this claim is ill-founded and this simplistic view is misleading\footnote{There is plenty of truth in the idiom,``truth is always simple, but simplicity is not always the truth.''}.   We do this by working out in  detail the three cases which had been studied before -- a free particle \cite{KP98},  an inverted harmonic oscillator \cite{GuthPi85} and the inflationary universe \cite{PS96} --  pointing out the exact places where illegitimate jumps were made, which prompted these claims and promulgated this erroneous view.

With a simple quantum harmonic oscillator example, using both the Wigner function and the wave function, and an operator Heisenberg equation,  we examine several commonly used criteria of classicalization, including the commonplace $\hbar\to 0$, and the somewhat more sophisticated  large $n$ approximation. On the other hand, for a free particle, and inverted oscillator, we focus on the late time, highly squeezed, limit when the system  is taken to behave classically by many authors. The squeezing under time evolution  turns a  quadrature  ellipse in phase space into a very narrow and elongated shape.  This is where many authors made a leap of faith and claim that the ellipse is like a  line, replace it by a delta function {in the expression of the corresponding Wigner function}, which perfects a trajectory in phase space and proclaim classicalization consummated. We  point out  this act is illegitimate, because if treated so,   negative eigenvalues and unphysical states arise.    

For the inflationary universe, we  follow  the evolution of the squeezed quantum  field, the inflaton {perturbation}, with nonequilibrium quantum field theory treatment {in terms of the Bogoliubov coefficients}, and demonstrated that entanglement persists between the pair of particles created. Entanglement being a uniquely quantum feature absent in classical physics,  this is an unequivocal evidence that the system does not turn classical even at late times under severe inflation.  An important criterion for all four examples we invoked is the Robertson-Schr\"odinger uncertainty relation (not the Schr\"odinger uncertainty as used in \cite{GuthPi85}), which is an {invariant} throughout the unitary evolution. Other criteria we have used include the non-commutativity of operators, boundedness of the Wigner function, and semi-positivity of the density matrix.   

The contrast with PS is even clearer:  while PS asserted that there is decoherence with a clever twist, we show that there is no decoherence, no twist. 


\subsection{Gravitational and quantum field perturbations -- some clarifying remarks}

Because quantum  cosmological perturbations involve both classical gravitational perturbations and quantum field fluctuations,  some clarification remarks may be needed on a few basic issues related to the role of quantum field fluctuations in cosmological perturbations. This subsection provides the background for, and can be read as a preamble of,  Sec.~\ref{S:ebjfkd}.

There are three parties involved in inflationary cosmological perturbations: i) classical gravitation theory,  based on general relativity, which governs cosmological evolution, ii) a quantum field, the inflaton, which drives the universe to inflationary expansion, and iii) their perturbations/fluctuations:  the scalar sector of gravitational perturbations is coupled to the quantum scalar field fluctuations, together governing the density contrasts which seed the  structures in the universe, like galaxies. 

 Classical theories of cosmological structure formation are based on gravitational perturbation theory \cite{Lif46,LifKha63,Haw66,Bardeen} where the density contrasts, the isocurvature perturbations, the vorticity and the primordial gravitational waves are described by the scalar, vector and tensor {perturbations} of the background spacetime.  The gravity sector based on general relativity  is classical throughout.  We will comment on tensor perturbations and graviton physics separately.  

  The scalar (inflaton) field $\hat \Phi ({\bm x}, t)$ is intrinsically quantum in nature.   Often a background field expansion $\hat \Phi ({\bm x}, t) = \bar \phi ({\bm x}, t)+ \delta \hat {\varphi} ({\bm x}, t)$ is performed,  where the background field $\bar \phi ({\bm x}, t)$ is a  mean field\footnote{The background field is often assumed to be classical, but this  needs to be proven rather than assumed to be automatically valid.  The $\bar \phi ({\bm x}, t)$ regarded as a  mean field keeps its quantum nature. (The difference between a mean field and a classical field is explained in, e.g., \cite{MQP1}.)  One can  show how readily the mean field is decohered by its quantum fluctuations, such as treated in \cite{CalHu95}.} and $\delta \hat {\varphi} ({\bm x}, t)$ are the quantum fluctuations (N.B. strictly speaking, quantum perturbations -- see Sec.~\ref{Sruherds} below). 
To get compact equations of motion  for the density contrasts,  mixed metric perturbations + quantum scalar field variables  are used, such as the gauge invariant Mukhanov-Sasaki variable.   Now, with a mixed variable coming from classical gravity and quantum field,  what do  quantum cosmological perturbations refer to,  and which variables are we targeting in their decoherence, or, which quantum variables become classical at late times -- or do they? 

\subsubsection{Which quantities in cosmological perturbations are intrinsically quantum?}
  
The mixed gravity + inflaton variables can come in many shades depending on which gauge one chooses to use in the gravity sector and the apportioned weight of each sector.  Regardless, the gravitational perturbations remain of classical origin. The scalar sector of the metric perturbations related to the Newtonian potential is a constraint, not a dynamical degree of freedom (the tensor modes are).  Its nature is determined by (or `slaved' \cite{AnaHuGauge,AnaHuGauge1} to) the matter source. In general relativity when the matter source is classical, this scalar sector of the metric perturbation is classical.  

In an extreme case, one may   conjure up situations where the quantum  fluctuations of the scalar field are made to vanish, such as ``choosing a  (co-moving) gauge which for scalar perturbations makes the velocity perturbation vanish.  For single field inflation,  this means that the time coordinate is defined so that at any given time the scalar field equals its unperturbed value" (one is riding up and down with the scalar field's fluctuations),  ``with all perturbations relegated to components of the metric'' (\cite{Weinberg}, Sec.~5.3D). This does not mean that gravity has become quantum, only that the scalar perturbations now acquire a quantum nature by virtue of the presence of the inflaton field.   When there is no inflaton, one returns to purely classical gravity.  The Newtonian force is slaved to the source which is classical. There is no way for the gravitational perturbations to become quantum in this way, and there are no decoherence issues.  

\subsubsection{Tensor perturbations: gravitational waves. Quantized tensor perturbations: gravitons}

The tensor sector of gravitational perturbations  are not linked to the quantum field which drives inflation and thus there is no issue of quantumness by proxy with the inflaton. They are intrinsically classical and carry gravity's dynamical (or propagating) degrees of freedom.   Primordial gravitational waves are described by the tensor sector of the gravitational metric perturbations. They have  been studied at the classical level since 1946 by Lifshitz and others.  One can consider quantizing the  linearized tensor perturbations, whence they become the gravitons, like photons for QED\footnote{Note gravitational waves are weak metric perturbations. Gravitons are  quantized short wavelength linear perturbations off of a smooth spacetime manifold,  in the nature of collective excitations.  Gravitons are governed by general relativity,  a low energy theory  for the  macroscopic structure of spacetime,  a far cry from quantum gravity, defined as  theories for the microscopic structures of spacetime functional at the Planck scale \cite{E/QG}}. Primordial gravitons created from the vacuum fluctuations in the early universe have been studied by many authors since the 70s \cite{Grishchuk74,ForPar77}. The two polarizations $(+, \times)$ each obey an equation of motion of the same form as a massless minimally-coupled scalar field.  The normal mode amplitudes of each polarization obey an equation of motion of the same form as that of a parametric oscillator with time-dependent frequency determined by the expansion of the universe, as studied here and earlier (e.g.,\cite{IntEnt} and references therein). Decoherence of primordial gravitons and decoherence due to gravitons are important current subjects  which we hope to return in conjunction with graviton detection \cite{PWZ,KST21,ChoHu21} and gravitational decoherence \cite{Bassi,Bao,AnaHu21} issues.  

\subsubsection{Perturbations: deterministic variables.  Fluctuations: stochastic variables}\label{Sruherds}
It is of theoretical significance to make the distinction between \textit{quantized linear perturbations},  which are believed to be the progenitors of galaxies and structures we see today,and \textit{vacuum fluctuations of a quantum field}, which engender spontaneous creation of particle pairs, a subject fundamental in quantum field theory in curved spacetime.  Note the former is a deterministic variable whereas the latter a stochastic variable.  What is customarily called quantum `fluctuations' of the inflaton field: the $\delta \varphi$ above should strictly speaking be called perturbations, because they are deterministic variable, obeying deterministic equations of motion    It is important to make this distinction especially when people try to replace quantum field-theoretical variables  by classical stochastic variables. 
The relation between a quantum variable and a classical stochastic variable is a nontrivial one. A lot depends on what constitutes the noise, how it is introduced and how it acts on a system. See, e.g., \cite{GelHar93}.   Some features of the former can be captured by the latter, but not all.  For Gaussian systems it is easier to bridge the two, but still there remain differences.  See discussions in, e.g.,  the last section of \cite{IntEnt}.

Fluctuations in (linear) quantum matter fields can be represented by the noise kernels (vacuum expectation values of the stress energy bitensor). When these fluctuations are included in addition to the expectation values of the stress-energy tensor (the mean)  as   sources driving the Einstein equation, they  induce  metric fluctuations (`spacetime foam'). There, fluctuations are of the main concern, and the Einstein-Langevin equation is the centerpiece of semiclassical stochastic gravity. 

\subsection{Model  Studies, Key Findings and Organization}\label{S:ribdfgd}

Of the four veins of decoherence studies described in more detail in the next section we shall  focus on one of these veins, the evolution of closed quantum systems.
Two representative work are, the 1985 paper of Guth and Pi (GP)~\cite{GuthPi85}  which contains great details, and the 1996 paper of Polarski and Starobinsky (PS)~\cite{PS96} which we mentioned earlier, together with subsequent joint papers with Kiefer along the same thread~\cite{KPS98}.  This vein does not require any  environmental field to decohere the inflaton perturbations, but  focuses on the late time behavior in the evolution of the inflaton field perturbations. Since both sets of papers use simple quantum mechanical models to illustrate their findings, the inverted harmonic oscillator (IHO) of GP, and the free particle of Kiefer et al,  we shall do the same, so direct comparisons can be made to see the differences. 

Before we delve into the details of the model systems studied  {in Sec.~\ref{S:dkhejre}}, we first use the simple harmonic oscillator model to point out relevant subtleties in taking the semi-classical limit. In particular, we address 1) whether/how the regions, in which the Wigner function assumes negative values, vary when the small $\hbar$ limit is taken, 2) the (in)compatibility between the different protocols of taking the semi-classical limits such as taking the large $n$ (excitation) limit vs the small $\hbar$ limit. This is especially pertinent to our subsequent analysis,  {warning against treating large squeezing as the classical limit}, and 3) offer a pedagogical derivation to show how classical dynamics in phase space can indeed emerge, but only upon taking the proper semi-classical limit.  

The model systems we investigate in this paper have a common feature that the ellipse in phase space formed by the dispersion of the canonical variables of the model system becomes exceedingly squashed in one quadrature but extremely stretched in another. {Prior authors made the observation that to leading order of the large squeezing parameter in their states, the ellipse reduces to a one-dimensional path in phase space. From this, they argue that an apparent classicality emerges from the quantum systems of these models. They also showed that the accompanying Wigner function is proportional to a delta function, and used this as a heuristic support for their claims.}  However, as we shall show in this paper,  their claims are invalid. It is dangerous to keep only the dominant contributions in treating the semiclassical limit. Doing so will have the following unfortunate consequences:
\begin{enumerate}
	\item Such a Wigner function does not correspond to a physical state
		\begin{enumerate}
			\item it violates the bound of the Wigner function when $\hbar\neq0$,
			\item if the system started in a pure state, the final state is no longer pure even though the evolution is unitary,
			\item the purity of the state is greater than unity,
			\item this implies that the corresponding density matrix has negative eigenvalues,
		\end{enumerate}
	\item The Robertson-Schr\"odinger uncertainty relation for a pure state is violated,
	\item The commutator of the canonical operator, like $\hat{x}$, at different times becomes commutative,
	\item The equal-time commutation relation of the canonical variables vanishes so the canonical pair commutes. 
\end{enumerate}
These behaviors are contradictory to our understanding of how a closed quantum system undergoes unitary evolution. These fallacies could have been avoided if   the sub-dominant contributions had been included in one's consideration.  Even though these subdominant terms are likely to be very small compared to the dominant ones, they are just what we need to keep things right. {Coherence,  which carries the quantum  essence of the closed system, resides in these sub-dominant contributions.  In addition, in the case of cosmological perturbations treated in Sec.~\ref{S:ebjfkd}, an irrefutable support that such a system remains quantum comes from the existence of quantum entanglement between the particle pairs produced in the process of parametric amplification due to the expansion of the universe.}

In Sec.~\ref{S:erjbd}, we give a short review of the study of decoherence and quantum to classical transition in cosmology,  in five stages of development, featuring four main veins of approach. This is the quantum backdrop necessary for the investigation of decoherence in inflationary cosmological perturbations. 
In Sec.{~\ref{S:bkgd}, we start  with an overview of the quantum mechanical tools used in the Gaussian dynamics of closed linear systems, and stress the unique role of the Robertson-Schr\"odinger uncertainty principle for  pure quantum states. We then turn to the relevant properties of the Wigner function in the context of the quantum-to-classical transition, and show that the aforementioned pitfalls are quite generic when the limiting cases are taken without mindful discretion. In Sec.~\ref{S:dkhejre} we first use the harmonic oscillator model to shed light on the elusive points in taking the semi-classical limits. Then we use the free particle and the inverted harmonic oscillator models by previous authors to pinpoint where their seemingly plausible assumptions lead to adverse pitfalls in drawing conclusions regarding the emergence of classicality in these closed quantum systems under unitary evolution, and show how judicious measures lead to correct conclusions. Sec.~\ref{S:ebjfkd} is dedicated to the quantum perturbations of the  inflaton field.  In addition to the features of  the quantum mechanical models studied in Sec.~\ref{S:bkgd} and~\ref{S:dkhejre}, a new feature pertaining to quantum fields arises, namely, quantum entanglement among particle pairs produced out of the field quanta by the expanding universe. This is an unmistakable signifier of the quantumness of the inflaton field perturbations which cannot be erased by the simplistic arguments used by prior authors.  We summarize the major findings of this paper in the Conclusion section.  In Appendix~\ref{S:ehhee} we show that the squeeze transformation does not modify the bound in the generalized uncertainty relation for the free, linear quantum scalar  field. In Appendix~\ref{S:nxkje} we show the proper limits to take in a harmonic oscillator model to reach the correct semiclassical limit.

\section{Decoherence in Cosmology: highlights of past four decades}\label{S:erjbd}

 Before one delves into a study of the decoherence for quantum inflationary perturbations it is useful to be conversant of the ways how decoherence is addressed in simpler settings in quantum mechanics and in more complex settings in quantum cosmology.  There are about four decades of work  on decoherence (counting from Zurek's early papers \cite{Zurek,Zurek1}) and many approaches with very different emphasis. In this section,  as a refresh,  we shall outline the four main veins of decoherence,  so beginning researchers can appreciate the complexity of the issues involved and become aware of the variety of methodologies used.  It can also serve as a coordinate system for experienced practitioners to compare notes, to  identify different set-ups, to define the issues they want to address and the approaches they wish to take.   Readers already familiar with this subject can skip over this section and proceed to Sec.~\ref{S:bkgd}.  

\subsection{Background (5 stages) and Methodology (4 veins)}

{\bf O. 1980s}.  The {\bf Theoretical Foundations} were laid down in the 80s.  A) {\it Environment-Induced Decoherence},  in the work of  Zurek \cite{Zurek} and  Joos \& Zeh \cite{JoosZeh},  aided by the 1983 Caldeira-Leggett master equation \cite{CalLeg83} for quantum Brownian motion for Markovian quantum processes. B) {\it Consistent histories} of Griffiths \cite{Griffiths} and Omnes \cite{Omnes}.

{\bf I. 1985-88},  the first period of work on decoherence in  {\bf Inflationary Cosmology},    we mention two relevant papers:  1) Guth and Pi 1985 (GP)    \cite{GuthPi85}  used an uncertainty relation to demarcate between quantum and classical.  We refer to this approach as the {\bf First Vein: closed quantum system} to this issue, the vein pursued in this paper.   2)  Starobinsky's 1986 stochastic inflation \cite{StoInf}
(see also Sakagami \cite{Saka88}) as a representative of the {\bf Second Vein:  closed with partition} (or with this symbol $(> | <)$ where a {\it non}interacting  scalar field is partitioned into a long wavelength ($<$,  division according to wavevector $\bm k$: low $\bm k$ refers to long wavelength modes) segment and a short wavelength segment ($>$). The latter is assumed to be a white noise  in a Langevin equation which drives inflation.   The long wavelength segment is assumed to be classical.  Outstanding issues in this model include: a) Does a sharp cutoff indeed generate white noise \cite{VilWin};  b) with a shifting partition in $\bm k$ space  a proper treatment requires quantum field theory of half space with time-dependent Hilbert spaces, which is amiss;  c) Decoherence of the long wavelength sector. We shall continue this discussion along the Fourth Vein in the same spirit but with interacting quantum fields.  

{\bf II. 1988-1992}.  {\bf Quantum Cosmology}, where decoherence is considered for the emergence of classical spacetimes. Decoherence in quantum cosmology was a major focus in the work of many authors in the late 80s and early 90s, notably,  Gell-Mann \& Hartle \cite{GelHar89,Hartle90}, Halliwell \cite{HallQC}, Habib \& Laflamme \cite{HabLaf},  Kiefer \cite{KieferQC}, Singh \& Padmanabhan \cite{SinPad89,Pad89,PadSin90} , Paz \& Sinha \cite{PazSin91,PazSin92} and others.   Many papers invoke the  Born-Oppenheimer  approximation, in light of the discrepancy between the massive gravity sector and the ligher matter field sector,  and introduce a  {\it WKB time} in the Wheeler-DeWitt equation \cite{WheDeW,WheelerQGD},  enabling the wavefunctions of the universe to enter the semiclassical realm.  Strictly speaking the assumption is not decoherence. It is more in line with the slow variables of van Hove in statistical mechanics. The division between fast-slow  variables, heavy-light masses, high -low energy sectors and  treating them differently is the beginning step in an open quantum systems approach.  Such a view is applied to the minisuperspace approximation in, e.g., \cite{mss,Kiefer87,HPSmss}. 

{\bf III. 1990-96}. 

A. The consistent histories vein of the 80s is continued in the {\bf Decoherent Histories} of Gell-Mann and Hartle \cite{GelHar89,Hartle90,GelHar93,Hartle93}.   Of particular interest is how these authors use the existent conditions  of cosmology as a closed system to construct a new interpretation of quantum physics,  and then apply it to understand the quantum mechanics of spacetime \cite{HartleClosed1,HartleClosed2}. Its value  goes beyond the decoherence issues, probing  deeper into the relations between quantum and gravitation. 

B. Likewise,  the {\bf Environment-induced Decoherence} program has seen significant developments with the derivation of a {\it nonMarkovian master equation} for quantum Brownian motion~\cite{HPZ92,HPZ93} and applications to decoherence of model quantum systems~\cite{PHZ93}. For example,  Zurek, Habib, \& Paz~\cite{ZHP93} explained why coherent state has the most `classical' features,    Hu \& Zhang~\cite{HuZha,HuZha1}   derived an uncertainty relation at finite temperatures and use it to understand the quantum-classical demarcation.  We shall invoke a generalized (Robertson-Schr\"odinger) uncertainty relation in this work.   In 1994 Hu \& Matacz~\cite{HM94} derived the HPZ equations for a parametric Brownian oscillator in a parametric oscillator bath -- parametric refers to oscillators with time-dependent frequencies in their normal modes.  This is  useful for treating squeezed states in quantum optics and for cosmology. As we shall see,  cosmological perturbations obey equations with a time-dependent effective mass, and cosmological particle creation can be seen as a manifestation of the quantum field  being squeezed~\cite{GriSid,HKM94,EntSqV,InfSqV} by the expansion of the universe.  With this connection one can investigate the entropy, decoherence and entanglement issues related to quantum cosmological perturbations. (For a brief description, see, e.g.,~\cite{IntEnt} and the references therein.)

C.  For  issues in {\bf inflationary  cosmology},  we mention three groups of papers in that period which exemplify the second vein and introduce two additional  veins in the approaches to cosmological decoherence. 
As  background on cosmological perturbations,  decoherence and entropy  issues, read the papers of Brandenberger, Laflamme, Mukhanov, Prokopec, Gasperini and Giovannini, et al~\cite{BLM,MFB,BMP,BMP1,GasGio,GasGio1}, and the recent mini-review~\cite{IntEntr} where many references can be found.  
 
{\bf Third Vein:  two-fields}: {\it nonMarkovian master equation with colored noise}. 1) The first serious study of cosmological decoherence based on nonMarkovian master equations for a system quantum field and a bath  quantum fields is in the 1992 paper of Hu, Paz \& Zhang (HPZ)~\cite{HPZdec}. These authors consider   two independent self-interacting $\phi^4$ scalar fields in de Sitter spacetime: $\lambda_\phi \phi^4$ depicting  the system, and $\lambda_\psi \psi^4$ depicting the bath,  and an interaction between them in the biquadratic form $\lambda_{\phi\psi}\phi^2\psi^2$. (Note that a system interacting with an environment with an interaction action of the bilinear form $\phi \psi $ such as used in Cornwall \& Bruisma~\cite{CorBru88} would not decohere\footnote{Despite the similarity in form with the bilinear $xq_n$ type of coupling between a quantum Brownian oscillator interacting with a bath of many modes,  when two fields are bilinearly coupling,  only one mode of the system field interacts with one mode of the bath field,  the physics is totally different. It is like two equal subsystems interacting. One would not see dissipation or decoherence,  the energy and phase information only pass from one to another back and forth.  A large number of modes in the bath is needed to see dissipation and decoherence in the system.}.) These authors used this model to address two basic issues in the theory of galaxy formation from the fluctuations of quantum fields: a) the nature and origin of noise and fluctuations and b) the conditions which need to be met for using a classical stochastic equation for their description.   Whether  the stochastic inflation proposal~\cite{StoInf} can fly depends critically on a satisfactory resolution of these two issues. 

On the first issue, HPZ derived the influence functional for a $\lambda \phi^4 $ field in a zero-temperature bath in de Sitter universe and obtained the correlators for the colored noises of vacuum fluctuations.  This exemplifies a new mechanism for colored noise generation which can  act as seeding for galaxy formation with non-Gaussian distributions.  For the second issue, HPZ presented a (functional) master equation for the inflaton field in de Sitter universe. By examining the form of the noise kernel they studied the decoherence of the long-wavelength sector and the conditions for it to behave classically.  The more general case of the system field and bath field interacting with the form $\lambda_{\phi\psi}f[\phi(x)]\psi^k$ was deal with by Zhang in his thesis work~\cite{ZhangPhD}, as reported in~\cite{Banff},  based on the functional perturbative methods they used for the study of nonlinear QBM~\cite{HPZ93}. 

{\bf  Fourth Vein:  nonlinear fields}  --  2a) {\it decoherence of the mean field by quantum fluctuations}. Instead of using a rather ad hoc splitting of a quantum field in stochastic inflation~\cite{StoInf}  into long and short wavelength segments, with the latter providing the noise which decoheres the former, Calzetta Hu  in 1995~\cite{CalHu95} treated a nonlinear field and examined the decoherence of the mean field by the interacting field's own quantum fluctuations, or that of other fields it interacts with. 
Note, in spirit, this shares with the first vein for a closed quantum system.  The quantum field by itself is  a closed system. The decomposition of  an interacting quantum field  into a mean field and the  fluctuation field is not enough to turn the mean field classical 
(which is commonly assumed in a background field expansion). One needs to specify the conditions when,  the ways how, and how likely a mean field will be decohered. Interaction with its quantum fluctuations is one way this may happen.    

2b) {\it Partitioning one interacting field into high and low frequency sectors}.   Starobinsky proposed this for a free field in his 1986 stochastic inflation model (See also Matacz~\cite{Matacz97}). We mentioned earlier the difficulties this model encounters as it is based on the partitioning of a free field, how the high frequency segment  acts on the low frequency segment is not clear, and a proper quantum field theory treatment is not easy. Instead, for an interacting quantum field with $(>|<)$ partition, there are no such conceptual issues. Technically it may seem more challenging, but there are well-developed methods to handle it, known as the  coarse-grained effective action~\cite{cgea,cgea1,CHM00}, the closed-time-path  or in-in~\cite{CTP,CTP1,CTP2,CTP3}  version of it is akin to the influence action~\cite{IF,CalLeg83,IF2}.  This was carried out neatly by Lombardo \& Mazzitelli~\cite{LomMaz96} and applied to cosmological decoherence by   Lombardo \& Lopez-Nacir~\cite{LomNac05}.  A natural  partition is the horizon scale,  in which case one can talk about entanglement entropy between the sub and super-horizon sectors. See, e..g,~\cite{Brahma}. 

We see that the conceptual and technical foundation for the study of cosmological decoherence were quite well established from 1982 to 1996.  They form the theoretical frameworks for continued investigations in the twenty five years following.

{\bf IV. 1996-2008.} Selected representative works include:

A. The 1996 paper of  Polarski \& Starobinsky (PS)~\cite{PS96}.  It belongs to the first vein, similar to Guth \& Pi, in that the authors assert that decoherence  comes by naturally without any environment assistance
(we shall refer to this as `intrinsic').  After Kiefer joined the collaboration~\cite{KPS98,KPS00,KLPS}   arguments were made more rigorous and connection with other work  derived from theories with  solid foundations was extended,  such as for the entropy of gravitons with quantum open squeezed system~\cite{KMH97}. 

B. Anderson et al~\cite{AMM05}.  This paper can be read as the continuation of~\cite{KME97}, which expounded the ideas of \cite{HuPav},  a good  representative of the First Vein. These authors showed that with respect to a second order adiabatic vacuum there is no decoherence in the setup of Polarski and Starobinsky. 

C.  Martineau \& Brandenberger  2005~\cite{MarBra05} investigated the gravitational backreaction of long wavelength (super-Hubble) scalar metric fluctuations on the perturbations themselves, due to the nonlinearity in the Einstein equations,   for a large class of inflationary models.  Martineau~\cite{Mar07} considered gravitational backreaction and interactions due to nonlinearities in the matter evolution equation in the $\phi^4$  chaotic inflation model.   Prokopec   and Rigopoulos~\cite{ProRig} used two decoupled massive fields to study the decoherence of curvature perturbations during inflation. 

D.  Campo and Parentani 2008~\cite{CamPar1} begins with an interacting quantum field, but quickly truncate the  correlation hierarchy at the Gaussian level, thus effectively acting like a free field.  Their result  of the entropy of cosmological perturbations in a closed system agrees with other approaches, such as in~\cite{KMH97,LCH10,IntEnt}. In the second paper~\cite{CamPar}, they use an open-quantum system approach, and obtain results which independent of the choice of gauge or basis, thus ``pointer states appear not to be relevant to the discussion", which seems counter to the claims in~\cite{KLPS} .

{\bf V. 2008-2020.}  A welcoming trend in the recent decade  is a broader recognition that the concepts and methods of open quantum systems, effective field theory and nonequilibrium quantum field theory  are essential to a more rigorous and thorough treatment which can provide a deeper and better understanding of cosmological quantum processes related to entropy, decoherence and entanglement. Representative papers are:   Boyanovsky 2015~\cite{Boyan}  employing techniques from nonequilibrium quantum field theory~\cite{NEqFT,cddn}. Working with two field models (Third Vein) he obtained a master equation from which he derived the corrections to the power spectrum, and drew implications for dark matter.  Hollowood and McDonald~\cite{HolMcD17} followed Boyanovksy's pathway and  studied the evolution of decoherence and the onset of classical stochastic behavior as modes exit the horizon.  Burgess et al  2015~\cite{Burgess},  using effective field theory   explored stochastic inflation via the Lindblad equation  for Markovian processes. Nelson 2016~\cite{Nelson} considered the effect of gravitational nonlinearity from expanding the Einstein-Hilbert action to third order in the  fluctuations and show that they provide a minimal mechanism for generating classical stochastic perturbations from inflation. This is in a similar spirit as~\cite{CalHu95} of  the  Fourth Vein. Allowing for changes in the  partitioning of the frequency sectors separating the system from its bath, Shandera et al~\cite{SAK18} derived the evolution equation for the density matrix of a UV- and IR-limited band of comoving momentum modes of the canonically normalized scalar degree of freedom in two examples of nearly de Sitter universes.  Finally,  Brahma et al~\cite{Brahma} studied the entanglement entropy in cosmology with the super- and sub-horizon partition. It also contains a comprehensive list of references for work on these topics up to 2020.

\section{Quantum states in a closed system do not turn classical}\label{S:bkgd}
In this section we ask the generic question whether and when a quantum state behaves classically in a closed system. We first give a short overview about the mathematical tools useful in the context of Gaussian states, discuss some important and often overlooked subtleties we will meet in the semi-classical limit.

Let $(x,p)$ be the pair of  canonical variables of a closed Gaussian system, where $p$ is the momentum conjugated to the position $x$. The system is assumed in a Gaussian state. In quantum mechanics, this canonical pair will be promoted to operators $(\hat{x},\hat{p})$.

\subsection{Heisenberg equation}
The Heisenberg picture offers insights into the non-commutativity of operators. The Heisenberg equations of operators offer a very intuitive way to investigate the quantum dynamics of linear (interacting) systems. For a Gaussian system, the evolution of its canonical-variable operators can be written as
\begin{align}
	\hat{x}(t)&=d_{1}(t)\,\hat{x}(0)+\frac{d_{2}(t)}{m}\,\hat{p}(0)\,,&\hat{p}(t)&=m\dot{d}_{1}(t)\,\hat{x}(0)+\dot{d}_{2}(t)\,\hat{p}(0)\,,
\end{align}
in terms of their initial values, $\hat{x}(0)$ and $\hat{p}(0)$ at the initial time $t=0$. The parameter $m$ denotes mass if the system is a linear (harmonic/inverted) oscillator, or it is set to unity if the system describes the modes of a linear field. Here $d_{1,2}(t)$ are a special set of homogeneous solutions to the equation of motion, called the fundamental solutions, satisfying
\begin{align}
	d_{1}(0)&=1\,,&\dot{d}_{1}(0)&=0\,,&d_{2}(0)&=0\,,&\dot{d}_{2}(0)&=1\,.
\end{align}
This allows us to readily write down various moments of the Gaussian system
\begin{align}\label{E:fgeusdfg}
	\langle\hat{x}(t)\rangle&=d_{1}(t)\,\langle\hat{x}(0)\rangle+\frac{d_{2}(t)}{m}\,\langle\hat{p}(0)\rangle\,,&\langle\hat{p}(t)\rangle&=m\dot{d}_{1}(t)\,\langle\hat{x}(0)\rangle+\dot{d}_{2}(t)\,\langle\hat{p}(0)\rangle\,,
\end{align}
and then
\begin{align}
	\langle\hat{x}^{2}(t)\rangle&=d_{1}^{2}(t)\,\langle\hat{x}^{2}(0)\rangle+\frac{2}{m}d_{1}(t)d_{2}(t)\,\frac{1}{2}\langle\bigl\{\hat{x}(0),\hat{p}(0)\bigr\}\rangle+\frac{d_{2}^{2}(t)}{m^{2}}\,\langle\hat{p}^{2}(0)\rangle\,,\\
	\langle\hat{p}^{2}(t)\rangle&=m^{2}\dot{d}_{1}^{2}(t)\,\langle\hat{x}^{2}(0)\rangle+2m\dot{d}_{1}(t)\dot{d}_{2}(t)\,\frac{1}{2}\langle\bigl\{\hat{x}(0),\hat{p}(0)\bigr\}\rangle+\dot{d}_{2}^{2}(t)\,\langle\hat{p}^{2}(0)\rangle\,,\label{E:gksjgs}\\
	\frac{1}{2}\langle\bigl\{\hat{x}(t),\hat{p}(t)\bigr\}\rangle&=md_{1}(t)\dot{d}_{1}(t)\,\langle\hat{x}^{2}(0)\rangle+\bigl[d_{1}(t)\dot{d}_{2}(t)+\dot{d}_{1}(t)d_{2}(t)\bigr]\,\frac{1}{2}\langle\bigl\{\hat{x}(0),\hat{p}(0)\bigr\}\rangle\notag\\
	&\qquad\qquad\qquad\qquad\qquad\qquad\qquad+\frac{1}{m}d_{2}(t)\dot{d}_{2}(t)\,\langle\hat{p}^{2}(0)\rangle\,.
\end{align}
The dispersions follow similar structure with, for example, $\langle\hat{x}^{2}(0)\rangle$ replaced by $\langle\Delta\hat{x}^{2}(0)\rangle$, where $\Delta\hat{x}(t)=\hat{x}(t)-\langle\hat{x}(t)\rangle$. We observer that
\begin{align}\label{E:eurbdfhd}
	\bigl[\hat{x}(t),\hat{x}(t')\bigr]=\frac{1}{m}\Bigl[d_{1}(t)d_{2}(t')-d_{1}(t')d_{2}(t)\Bigr]\times\bigl[\hat{x}(0),\hat{p}(0)\bigr]\,.
\end{align}
In general \eqref{E:eurbdfhd} does not vanish. That is, the operator $\hat{x}$ at different times does not commute. Similarly we can show the equal-time canonical commutation relation is obeyed for all times 
\begin{equation}
	\bigl[\hat{x}(t),\hat{p}(t)\bigr]=\Bigl[d_{1}(t)\dot{d}_{2}(t)-\dot{d}_{1}(t)d_{2}(t)\Bigr]\times\bigl[\hat{x}(0),\hat{p}(0)\bigr]=\bigl[\hat{x}(0),\hat{p}(0)\bigr]=i\,,
\end{equation}
due to the Wronskian conditions of the fundamental solutions.

\subsection{Gaussian pure state}
Now we go to the Schr\"odinger picture, and the dynamical evolution of the linear system is fully accounted for by the wave function. Consider a general time-dependent Gaussian pure state
\begin{equation}
	\psi(x,t)=\mathsf{a}(t)\,\exp\Bigl[-\mathsf{b}(t)\,x^{2}+i\,\mathsf{c}(t)\,x\Bigr]\,.
\end{equation}
The normalization condition enables us to write the wavefunction into the form
\begin{equation}\label{E:etuedgfb}
	\psi(x,t)=\Bigl(\frac{2\operatorname{Re}\mathsf{b}}{\pi}\Bigr)^{\frac{1}{4}}\frac{\mathsf{a}}{\lvert\mathsf{a}\rvert}\,\exp\Bigl[-\mathsf{b}\,x^{2}+i\,\mathsf{c}\,x-\frac{(\operatorname{Im}\mathsf{c})^{2}}{4\operatorname{Re}\mathsf{b}}\Bigr]\,.
\end{equation}
We see that $\mathsf{a}(t)$ only contributes to an overall spatially independent, but time-dependent phase, so it will not enter the calculations of the covariance matrix elements. We then find
\begin{align}
	X&=\langle\hat{x}\rangle=-\frac{\operatorname{Im}\mathsf{c}}{2\operatorname{Re}\mathsf{b}}\,,&\langle\hat{x}^{2}\rangle&=\frac{1}{4\operatorname{Re}\mathsf{}b}+\frac{(\operatorname{Im}\mathsf{c})^{2}}{4(\operatorname{Re}\mathsf{b})^{2}}\,,&b&=\langle\Delta\hat{x}^{2}\rangle=\frac{1}{4\operatorname{Re}\mathsf{b}}\,,\\
	P&=\langle\hat{p}\rangle=\operatorname{Re}\mathsf{c}+\frac{\operatorname{Im}\mathsf{b}\operatorname{Im}\mathsf{c}}{\operatorname{Re}\mathsf{b}}\,,&\langle\hat{p}^{2}\rangle&=\frac{\lvert\mathsf{b}\rvert^{2}}{\operatorname{Re}\mathsf{b}}+\Bigl(\operatorname{Re}\mathsf{c}+\frac{\operatorname{Im}\mathsf{b}\operatorname{Im}\mathsf{c}}{\operatorname{Re}\mathsf{b}}\Bigr)^{2}\,,&a&=\langle\Delta\hat{p}^{2}\rangle=\frac{\lvert\mathsf{b}\rvert^{2}}{\operatorname{Re}\mathsf{b}}\,,
\end{align}
and
\begin{align}
	\frac{1}{2}\langle\bigl\{\hat{x},\hat{p}\bigr\}\rangle&=-\frac{\operatorname{Im}\mathsf{b}(\operatorname{Im}\mathsf{c})^{2}}{2(\operatorname{Re}\mathsf{b})^{2}}-\frac{\operatorname{Im}\mathsf{b}}{2\operatorname{Re}\mathsf{b}}-\frac{\operatorname{Re}c\operatorname{Im}\mathsf{c}}{2\operatorname{Re}\mathsf{b}}\,,&c&=\frac{1}{2}\langle\bigl\{\Delta\hat{x},\Delta\hat{p}\bigr\}\rangle=-\frac{\operatorname{Im}\mathsf{b}}{2\operatorname{Re}\mathsf{b}}\,.
\end{align}
The expressions of the coefficients $\mathsf{a}$, $\mathsf{b}$ and $\mathsf{c}$ will be determined by the Schr\"odinger equation. Note that the cross correlation between the canonical variables does not vanish unless $\operatorname{Im}\mathsf{b}=0$.

If we use the pure state \eqref{E:etuedgfb} to construct the density matrix elements, we have
\begin{align}\label{E:dkgbsera}
	\rho(x,x';t)=\Bigl(\frac{2\operatorname{Re}\mathsf{b}}{\pi}\Bigr)^{\frac{1}{2}}\,\exp\Bigl[-\mathsf{b}\,x^{2}-\mathsf{b}^{*}\,x'^{2}+i\,\mathsf{c}\,x-i\,\mathsf{c}^{*}\,x'-\frac{(\operatorname{Im}\mathsf{c})^{2}}{2\operatorname{Re}\mathsf{b}}\Bigr]\,.
\end{align}
Note that there is no $xx'$ term in the exponent. However this condition may not serve as a criterion that the state is pure. This is basis  dependent. For example, if we change to the $(\Sigma,\Delta)$ bases, Eq.~\eqref{E:dkgbsera} becomes
\begin{align}
	&\quad\rho(\Sigma,\Delta;t)\\
	&=\Bigl(\frac{2\operatorname{Re}\mathsf{b}}{\pi}\Bigr)^{\frac{1}{2}}\,\exp\Bigl[-2\operatorname{Re}\mathsf{b}\,\Sigma^{2}-\frac{1}{2}\operatorname{Re}\mathsf{b}\,\Delta^{2}-i\,2\operatorname{Im}\mathsf{b}\,\Sigma\Delta-2\operatorname{Im}\mathsf{c}\,\Sigma+i\,\operatorname{Re}\mathsf{c}\,\Delta-\frac{(\operatorname{Im}\mathsf{c})^{2}}{2\operatorname{Re}\mathsf{b}}\Bigr]\,.\notag
\end{align}
The coefficient of the $\Sigma\Delta$ term is nonzero, but obviously it still describes a pure state.

\subsection{Wigner function and density matrix elements}
The covariance matrix $\mathbf{C}$ turns out to be an convenient building blocks of the Gaussian system, and for a one-dimensional system it is defined as
\begin{align}\label{E:iertnsdbf}
	\mathbf{C}&=\begin{pmatrix}b&c\\c&a\end{pmatrix}=\langle\hat{\mathbf{R}}\cdot\hat{\mathbf{R}}^{T}\rangle\,,	&\mathbf{R}&=\begin{pmatrix}x\\p\end{pmatrix}\,.
\end{align}
Here $\hat{\mathbf{R}}$ is the operator counterpart of $\mathbf{R}$, and we have assumed that $\langle\hat{\mathbf{R}}\rangle=0$. If not, we simply replace $\hat{\mathbf{R}}$ in \eqref{E:iertnsdbf} by $\hat{\mathbf{R}}-\langle\hat{\mathbf{R}}\rangle$. The elements of the covariance matrix have specific meanings. Since they are
\begin{align}
	b&=\langle\Delta\hat{x}^{2}\rangle\,,&a&=\langle\Delta\hat{p}^{2}\rangle\,,&c&=\frac{1}{2}\langle\bigl\{\Delta\hat{x},\Delta\hat{p}\bigr\}\rangle\,,
\end{align}
we see that $b$ gives the position dispersion, $a$ the momentum dispersion, and $c$ is the correlation between the $x$ and $p$ quadratures. In general, they are time-dependent functions and in general $c\neq0$.  In terms of these elements, the Robertson-Schr\"odinger uncertainty relation is
\begin{equation}\label{E:dgbksbf}
	ab-c^{2}\geq\frac{\hbar^{2}}{4}\,.
\end{equation}
Hereafter we will choose the units such that $c=\hbar=1$, but will put back $\hbar$ if necessary. The unitary evolution of the quantum system will not change the value of the lefthand side of \eqref{E:dgbksbf}.

The density matrix elements of a Gaussian state takes the form
\begin{equation}\label{E:rjghdgjfg}
	\langle x\vert\hat{\rho}(t)\vert x'\rangle=\rho(x,x';t)=\frac{1}{\sqrt{2\pi b}}\,\exp\Bigl[-\frac{1}{2b}\,\Sigma^{2}+i\,\frac{c}{b}\,\Sigma\Delta-\frac{ab-c^{2}}{2b}\,\Delta^{2}\Bigr]\,,
\end{equation}
with the simplification that the mean position and momentum are zero, and
\begin{align}
	\Sigma&=\frac{x+x'}{2}\,,&\Delta&=x-x\,,&&\Rightarrow&x&=\Sigma+\frac{\Delta}{2}\,,&x'&=\Sigma-\frac{\Delta}{2}\,.
\end{align}
The variable $\Delta$ gives a measure regarding the width of the off-diagonal elements perpendicular to the diagonal. In the absence of $c$, the parameter $a^{-1}$ gives the width of the off-diagonal elements. However when $c\neq0$, the interpretation becomes less transparent,
\begin{equation}
	\rho(x,x';t)=\frac{1}{\sqrt{2\pi b}}\,\exp\Bigl[-\frac{1}{2b}\,\bigl(\Sigma^{2}-i\,c\,\Delta\bigr)-\frac{a}{2}\,\Delta^{2}\Bigr]\,.
\end{equation}
The variable $a^{-1}$ still gives the width of the spread along the $\Delta$ direction, but the other orthogonal quadrature veers off the diagonal to $\Sigma-ic\,\Delta$, whose physical meaning is not transparent in this picture but will be better seen in terms of the Wigner function. Often we are interested in a quantity call purity, defined by $\langle x\vert\hat{\rho}^{2}(t)\vert x'\rangle$. From \eqref{E:rjghdgjfg}, we find
\begin{align}
	\langle x\vert\hat{\rho}^{2}(t)\vert x'\rangle&=\int\!dz\;\rho(x,z;t)\rho(z,x';t)\label{E:qnksfjer}\\
	&=\frac{1}{\sqrt{4\pi b(ab-c^{2}+1/4)}}\,\exp\biggl\{-\frac{4(ab-c^{2})}{4b(ab-c^{2}+1/4)}\,\Sigma^{2}+i\,\frac{c}{b}\,\Sigma\Delta-\frac{ab-c^{2}+1/4}{4b}\,\Delta^{2}\biggr\}\,.\notag
\end{align}
Thus when $ab-c^{2}=1/4$, we have $\langle x\vert\hat{\rho}^{2}(t)\vert x'\rangle=\langle x\vert\hat{\rho}(t)\vert x'\rangle$. It indicates that the density matrix describes a pure state\footnote{For a pure state $\lvert\psi\rangle$, the density matrix operator is $\hat{\rho}_{\psi}=\lvert\psi\rangle\langle\psi\rvert$. We thus have $\hat{\rho}_{\psi}^{2}=\hat{\rho}_{\psi}$. In addition, any pure Gaussian state can be reached from the vacuum state by a suitable unitary transformation. Since the vacuum has minimal uncertainty, that is, $ab-c^{2}=1/4$ and since the unitary transformation preserve the Robertson-Schr\"odinger uncertainty principle, the resulting pure Gaussian state then has $ab-c^{2}=1/4$. }. The trace of \eqref{E:qnksfjer} gives
\begin{equation}
	\operatorname{Tr}\hat{\rho}^{2}(t)=\frac{1}{2\sqrt{ab-c^{2}}}\,.
\end{equation}
For a pure state, we find $\operatorname{Tr}\hat{\rho}^{2}(t)=1$, while for a mixed state, we have $ab-c^{2}>1/4$, so we find $\operatorname{Tr}\hat{\rho}^{2}(t)<1$. This mixed state can be the reduced density matrix of a bi-partite pure entangled state.

The Wigner function $\mathcal{W}(x,p;t)$ of a Gaussian system is given by
\begin{align}\label{E:fkfjbsf}
	\mathcal{W}(x,p;t)&=\frac{1}{2\pi}\int\!d\Delta\;e^{-ip\Delta}\rho(\Sigma+\frac{\Delta}{2},\Sigma-\frac{\Delta}{2};t)=\frac{1}{2\pi\sqrt{\det\mathbf{C}}}\,\exp\Bigl[-\frac{1}{2}\,\mathbf{R}^{T}\cdot\mathbf{C}^{-1}\cdot\mathbf{R}\Bigr]\,,
\end{align}
with
\begin{align}
	\mathbf{C}^{-1}&=\frac{1}{\det\mathbf{C}}\begin{pmatrix}a&-c\\-c&b\end{pmatrix}\,.
\end{align}
It can be shown that
\begin{align}
	\langle\hat{x}^{2}\rangle&=\int\!dx\!\int\!dp\;x^{2}\,\mathcal{W}(x,p;t)=b\,,&\langle\hat{p}^{2}\rangle&=\int\!dx\!\int\!dp\;p^{2}\,\mathcal{W}(x,p;t)=a\,,\\
	\frac{1}{2}\langle\bigl\{\hat{x},\hat{p}\bigr\}\rangle&=\int\!dx\!\int\!dp\;xp\,\mathcal{W}(x,p;t)=c\,.\label{E:dfsjf}
\end{align}
In particular we note that \eqref{E:dfsjf} shows a correspondence between operator in Weyl ordering and its classical expression. The Wigner function provides an alternative formulation of quantum mechanics in terms of the phase space variables. In \eqref{E:fkfjbsf}, the Wigner function is positive definite in phase space. Thus it is often chosen as the candidate of a classical probability distribution, according to this observation. In fact, in general, it is not a positive definite, and can have negative values over regions, the area of each of which is of order $\hbar$ in a two-dimensional phase space~\cite{HO84,ML86,CF14}. We will provide a few examples at the end of this section.

When we write \eqref{E:fkfjbsf} explicitly in terms of covariance matrix elements,
\begin{equation}\label{E:fgsdewie}
	\mathcal{W}(x,p;t)=\frac{1}{2\pi\sqrt{ab-c^{2}}}\,\exp\Bigl[-\frac{a}{2(ab-c^{2})}\,x^{2}+\frac{c}{(ab-c^{2})}\,xp-\frac{b}{2(ab-c^{2})}\,p^{2}\Bigr]\,,
\end{equation}
we observe that if $c=0$, the exponent describes an ellipse whose semi-axes are $x$ and $p$ axes, and have lengths proportional to $\sqrt{b}=\sqrt{\langle\Delta\hat{x}^{2}\rangle}$ and $\sqrt{a}=\sqrt{\langle\Delta\hat{p}^{2}\rangle}$. Thus the $c\neq0$ case corresponds to a rotated ellipse.

When we investigate the distortion of quadratures, that is, squeezing or stretching, due to evolution, it is not easy to see the extent of distortion when $c\neq0$ because $a$ and $b$ do not tell the lengths of semi-axes. It proves convenient to co-rotate with the quadrature ellipse, that is, using the axes defined by the eigenvectors of the covariance matrix elements. The eigenvalues of the covariance matrix elements are given by
\begin{align}
	\lambda_{\pm}&=\frac{1}{2}\Bigl[\bigl(a+b\bigr)\pm\sqrt{\bigl(a-b\bigr)^{2}+4c^{2}}\Bigr]=\frac{1}{2}\Bigl[\bigl(a+b\bigr)\pm\sqrt{\bigl(a+b\bigr)^{2}-1}\Bigr]\,,\label{E:fgkjerd}
\end{align}
if $ab-c^{2}=1/4$, and the eigenvectors are
\begin{align}\label{E:jgskdfsd}
	\mathbf{v}_{-}&=\begin{pmatrix}1&+z\end{pmatrix}\,,&\mathbf{v}_{+}&=\begin{pmatrix}-z&1\end{pmatrix}\,,&&\text{and}&z&=\frac{\bigl(a-b\bigr)-\sqrt{\bigl(a-b\bigr)^{2}+4c^{2}}}{2c}\,.
\end{align}
Note that since the elements of the covariance matrix do not carry the same dimension, it is customary to append appropriate dimensional parameters to make expressions in \eqref{E:fgkjerd} and \eqref{E:jgskdfsd} dimensionally consistent.

The covariance matrix can be diagonalized by the matrix formed from the eigenvectors
\begin{align}
	\mathbf{M}&=\frac{1}{\sqrt{1+z^{2}}}\begin{pmatrix}1&-z\\+z &1\end{pmatrix}\,,&&\text{or}&\mathbf{M}^{T}&=\frac{1}{\sqrt{1+z^{2}}}\begin{pmatrix}1&+z\\-z &1\end{pmatrix}\,,
\end{align}
such that
\begin{equation}
	\mathbf{D}=\mathbf{M}^{T}\cdot\mathbf{C}\cdot\mathbf{M}=\begin{pmatrix}\lambda_{-}&0\\0 &\lambda_{+}\end{pmatrix}\,.
\end{equation}
It means the covariance matrix formed by a new canonical operator pair, $\hat{\bm{\mathsf{R}}}=\mathbf{M}^{T}\cdot\hat{\mathbf{R}}$, 
\begin{align}
	\hat{\mathsf{x}}&=\frac{1}{\sqrt{1+z^{2}}}\,\hat{x}+\frac{z}{\sqrt{1+z^{2}}}\,\hat{p}\,,&\hat{\mathsf{p}}&=-\frac{z}{\sqrt{1+z^{2}}}\,\hat{x}+\frac{1}{\sqrt{1+z^{2}}}\,\hat{p}\,,
\end{align}
is automatically diagonal, i.e., $c'=0$. By means of $\bm{\mathsf{R}}$, the Wigner function \eqref{E:fkfjbsf} becomes
\begin{align}
	\mathcal{W}(x,p;t)=\mathcal{W}(\mathsf{x},\mathsf{p};t)&=\frac{1}{2\pi\sqrt{a'b'}}\,\exp\Bigl[-\frac{\mathsf{x}^{2}}{2b'}-\frac{\mathsf{p}^{2}}{2a'}\Bigr]\,,\label{E:dkfghdf}
\end{align}
with
\begin{align}
	a'&=\langle\Delta\hat{\mathsf{p}}^{2}\rangle\,,&b'&=\langle\Delta\hat{\mathsf{x}}^{2}\rangle\,,&c'&=0\,.
\end{align}
The new canonical operator pair $\hat{\bm{\mathsf{R}}}$ is rotated from $\hat{\mathbf{R}}$ by an angle $\varphi$
\begin{equation}\label{E:dgsierie}
	\varphi=\tan^{-1}z\,.
\end{equation}
in phase space such that $[\hat{\mathsf{x}},\hat{\mathsf{p}}]=[\hat{x},\hat{p}]=i$. Thus $b'$ and $a'$ give us information about the semi-axes of the ellipse, and in turns, the extent of squeezing and stretching of the quadrature ellipse during the evolution.

\begin{figure}
\centering
    \scalebox{0.5}{\includegraphics{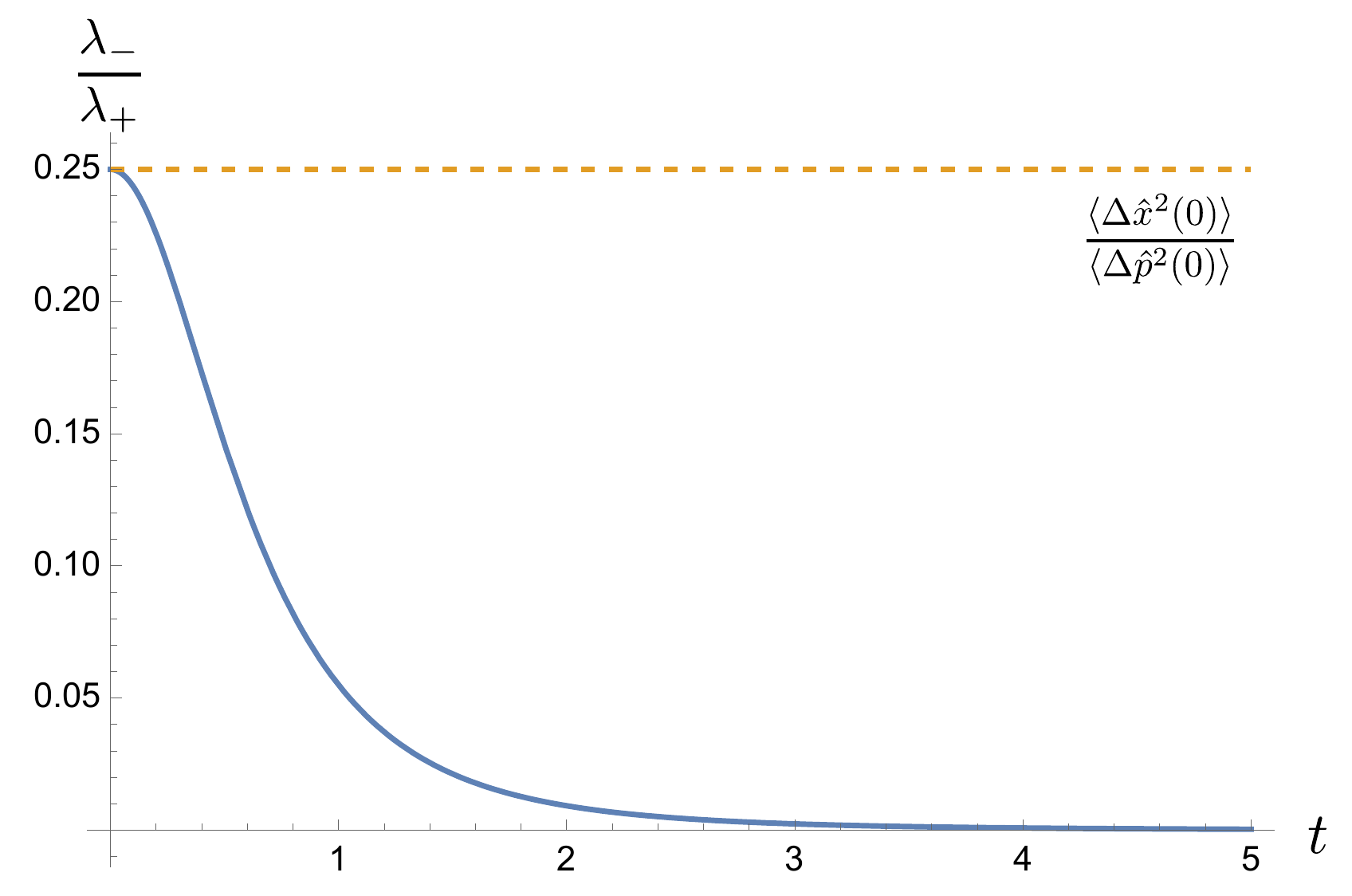}}
    \caption{The time variation of the ratio of the eigenvalues of the covariance matrix. it shows the squeezing and stretching of the quadratures. In this case, the position dispersion is highly squeezed but the momentum is stretched. For a free particle of mass $m$ and initial position dispersion $\sigma$, the ratio is asymptotically given by $\frac{4m^{4}\sigma^{4}}{t^{4}}$, and for the inverted harmonic oscillator, it behaves like $\frac{\lambda_{-}}{\lambda_{+}}=\frac{16\beta^{2}\sin^{4}2\theta}{[4\beta^{2}+1+(4\beta^{2}-1)\cos2\theta]^{2}}\,e^{-4\omega t}$, where the notations are explained in Sec.~\ref{S:dkhejre}.}\label{Fi:eigenratio}
\end{figure}

Let us take a special case $b'\to0$. We can use the formula
\begin{equation}\label{E:cnxlkjd}
	\delta(x)=\lim_{\epsilon\to0}\frac{1}{\sqrt{2\pi\epsilon}}\,e^{-\frac{x^{2}}{2\epsilon}}\,,
\end{equation}
to write the Wigner function into
\begin{equation}\label{E:nfskjds}
	\mathcal{W}(\mathsf{x},\mathsf{p};t)=\frac{1}{\sqrt{2\pi a'}}\,\exp\Bigl[-\frac{(\mathsf{p}-P)^{2}}{2a'}\Bigr]\,\delta(\mathsf{x}-X)\,,
\end{equation}
where we have put back the mean values $\langle\hat{\bm{R}}\rangle=(X,\,P)^{T}$. In this limit we find
\begin{equation}\label{E:woes}
	\lim_{b'\to0}\frac{\lambda_{-}}{\lambda_{+}}=0\,.
\end{equation}
If we take the limit according to \eqref{E:cnxlkjd} literally, then \eqref{E:woes} describes an ellipse which is extremely squeezed in one quadrature but extremely stretched in another. In particular the Wigner function \eqref{E:nfskjds} has an extremely sharp peak about $\mathsf{x}=X(t)$ and thus is often interpreted as a (quasi-)probability distribution along the one-dimensional path defined by the delta function, instead of over two-dimensional phase space. It is often claimed that in this case the Wigner function gives a classical probability description of a Gaussian state along a well-defined classical trajectory. Nonetheless, it does not meet our expectation that in the (semi-)classical limit, the system, averagely speaking, should follow the trajectory described by the mean position {$X(t)$, the expectation value of the canonical coordinate operator $\hat{X}$. The classical trajectory in phase space is a straight line parallel to the $x$ axis, different from the line defined by one of the semi-axes of the rotated ellipse, inferred by \eqref{E:nfskjds}}.

It is also interesting to observe that if we use the Wigner function \eqref{E:dkfghdf} to construct the density matrix elements, we find  
\begin{equation}\label{E:fgjhthsd}
	\rho(\Sigma',\Delta';t)=\frac{1}{\sqrt{2\pi b'}}\,\exp\Bigl[-\frac{1}{2b'}\,\Sigma'^{2}-\frac{a'}{2}\,\Delta'^{2}\Bigr]=\frac{1}{\sqrt{2\pi b'}}\,\exp\Bigl[-\frac{1}{2b'}\Bigl(\mathsf{x}^{2}+\mathsf{x}'^{2}\Bigr)\Bigr]\,.
\end{equation}
The limit $b'\to0$ implies that $a'\to\infty$ or $a'^{-1}\to0$. We then find that the density matrix elements in terms of these rotated variables $\Sigma'=(\mathsf{x}+\mathsf{x}')/2$ and $\Delta'=\mathsf{x}-\mathsf{x}'$ depict a highly localized, delta-function-like packet on the $\Sigma'$--$\Delta'$ plane or $\mathsf{x}$--$\mathsf{x}'$ plane, 
\begin{equation}\label{E:fgneris}
	\lim_{b'\to0}\frac{1}{\sqrt{2\pi b'}}\,\exp\Bigl[-\frac{1}{2b'}\Bigl(\mathsf{x}^{2}+\mathsf{x}'^{2}\Bigr)\Bigr]=\delta(\sqrt{\mathsf{x}^{2}+\mathsf{x}'^{2}})\,,
\end{equation}
even though the cross section of the quadrature profile in phase space is still an ellipse with the same area $\pi a'b'=\pi/4$, a consequence of invariance of the symplectic eigenvalues of the covariance matrix. This does not imply that the wavepacket is also localized in the $x$--$x'$ plane because in the $b'\to0$ limit the dispersion $\langle\Delta\mathsf{p}^{2}\rangle$ is essentially infinite. Note that so far we do not take $\hbar\to0$, and the limit $b'\to0$ is purely dynamical. Thus, obtaining a result like \eqref{E:fgneris} sounds odd. In fact, it has been shown~\cite{Ba58} that the Wigner function is bounded for a finite value of $\hbar$
\begin{equation}\label{E:fdgkdf}
	-\frac{2}{2\pi\hbar}\leq W(\mathsf{x},\mathsf{p})\leq+\frac{2}{2\pi\hbar}\,,
\end{equation}
so using \eqref{E:cnxlkjd} too literally introduces the artefacts to the Wigner function \eqref{E:nfskjds}, and makes it violate the bounds.

Another observation is made in~\cite{EF98}. The Gaussian Winger function~\eqref{E:fgsdewie} can be written as
\begin{align}\label{E:dgbsjkdhf}
	W(x,p;t)=\frac{1}{2\pi\sqrt{ab-c^{2}}}\exp\Bigl[-\frac{b}{2(ab-c^{2})}\Bigl(p-\frac{c}{b}\,x\Bigr)^{2}\Bigr]\,\exp\Bigl[-\frac{1}{2b}\,x^{2}\Bigr]\,.
\end{align}
In the limit $b\to\infty$,  we obtain  
\begin{equation}\label{E:ghdisue}
	\lim_{b\to\infty}W(x,p;t)=\frac{1}{\sqrt{2\pi b}}\,\delta(p-\frac{c}{b}\,x)\,\exp\Bigl[-\frac{1}{2b}\,x^{2}\Bigr]\,,
\end{equation}
with the help of \eqref{E:cnxlkjd}. This also shows the violation of \eqref{E:fdgkdf}, and if we compute the density matrix elements, denoted by $\varrho(x,x';t)$, from \eqref{E:ghdisue}, then we find
\begin{equation}\label{E:fgkfksd}
	\varrho(x,x';t)=\rho(x,x';t)\,\exp\Bigl[\frac{ab-c^{2}}{2b}\bigl(x-x'\bigr)^{2}\Bigr]\,,
\end{equation}
where $\rho(x,x';t)$ is given in \eqref{E:rjghdgjfg}. Three observations~\cite{EF98} are made on \eqref{E:fgkfksd}: a) the density matrix elements do not satisfy
\begin{equation}
	\int\!dx''\;\varrho(x,x'';t)\varrho(x'',x';t)=\varrho(x,x';t)\,,
\end{equation}
for $ab-c^{2}=1/4$. That is, $\varrho(x,x';t)$ does not describe a pure state, and not only that, b) the purity diverges
\begin{equation}
	\int\!dxdx'\;\varrho^{2}(x,x';t)=\infty\,,
\end{equation}
in contradiction to the fact that for a generic quantum state $\hat{\rho}$, its purity satisfies the bound
\begin{equation}
	\operatorname{Tr}\hat{\rho}^{2}\leq1\,.
\end{equation}
The purity of the pure state is equal to 1, while the mixed state has purity less than unity. Recall that we do not take $\hbar\to0$ yet. c) It is further argued in~\cite{EF98} that since
\begin{align}
	\operatorname{Tr}\hat{\varrho}&=1\,,&\operatorname{Tr}\hat{\varrho}^{2}&=\infty\,,
\end{align}
some of the eigenvalues of $\hat{\varrho}$ must be negative. It renders the von Neumann entropy associated with $\hat{\varrho}$ ill defined. Thus the state corresponds to the form of the Wigner function in \eqref{E:ghdisue} is unphysical. This example points out the subtleties in treating the limiting form of the Wigner function.

\section{Quantum Mechanical Examples}\label{S:dkhejre}
The Planck constant $\hbar$, a hallmark of quantum physics, does not reside in the classical descriptions of physical, chemical or the biological processes, so the limit $\hbar\to0$ in the formalism offers an unambiguous reduction from the quantum regime to the classical regime. However, in the operational sense, since $\hbar$ is a constant, taking this limit is not practically useful. We often turn to other parameters that are tunable and may be qualified for describing the quantum to classical transition of the system of our interest. One relevant to the discussion in this paper is the large $n$ limit, where $n$ can be the number of the constituents in the system, or represents the highly excited state of the system, such as an Rydberg atom~\cite{KL81}. This limit seems consistent with our mundane experience that a macroscopic system behaves classically. However here, as an appetizer, we will first use a simple example of quantum harmonic oscillator to illustrate the inequivalence of two limits.

\subsection{Harmonic Oscillator: semiclassical limit}  
\begin{figure}
\centering
    \scalebox{0.45}{\includegraphics{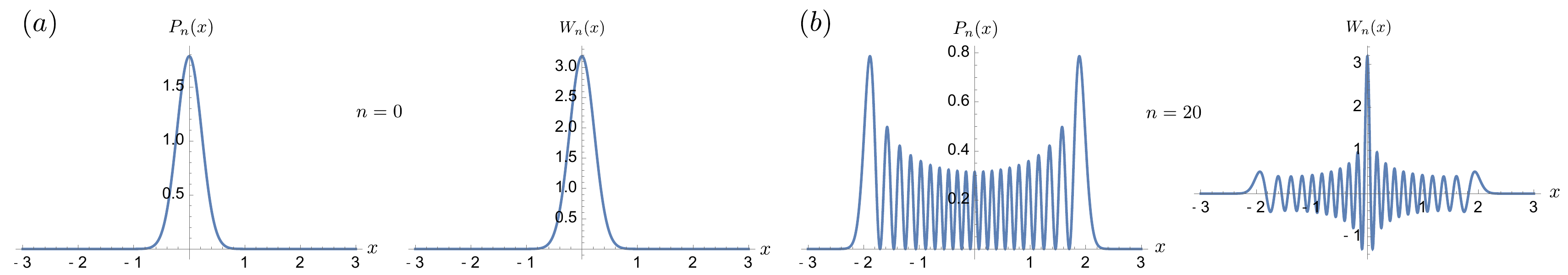}}
    \caption{(a) shows the probability distribution $P_{0}(x)$ and the Wigner function $W_{0}(x,p=0)$ of the ground state of a harmonic oscillator. Both are positive definite. (b) For the excited state $n=20$. It is clearly seen that the Wigner function becomes indefinite in sign.}\label{Fi:harmonOs-1}
\end{figure}

Consider the excited states of the harmonic oscillator, whose Hamiltonian is given by
\begin{equation}
	\mathcal{H}=\frac{p^{2}}{2m}+\frac{m\omega^{2}}{2}x^{2}\,,
\end{equation}
in which $m$ is the mass and $\omega$ is the oscillating frequency. The wavefunction of the $n^{\text{th}}$ excited state is 
\begin{align}
	\psi_{n}(x)&=\frac{1}{\sqrt{2^{n}n!}}\biggl(\frac{\alpha^{2}}{\pi}\biggr)^{\frac{1}{4}}e^{-\frac{\alpha^{2}x^{2}}{2}}H_{n}(\alpha x)\,,&\alpha^{2}&=\frac{m\omega}{\hbar}\,.
\end{align}
The probability density $P_{n}(x)$ of finding the oscillator in this excited state is
\begin{align}
	P_{n}(x)=\lvert\psi_{n}(x)\rvert^{2}=\frac{1}{2^{n}n!}\biggl(\frac{\alpha^{2}}{\pi}\biggr)^{\frac{1}{2}}e^{-\alpha^{2}x^{2}}\Bigl[H_{n}(\alpha x)\Bigr]{}^{2}\,,
\end{align}
where $H_{n}$ is the Hermite polynomial of the $n^{\text{th}}$ order. The probability distribution $P_{n}$ is always non-negative. In the classically allowed region, the probability function is oscillatory and the number of nodes is equal to $n$, but into the classically forbidden region, the probability exponentially decays, as shown in Fig.~\ref{Fi:harmonOs-1}.

The Wigner function of $n^{\text{th}}$ excited state of the harmonic oscillator is given by
\begin{align}
	W_{n}(x,p)&=\frac{1}{2\pi\hbar}\frac{1}{2^{n}n!}\biggl(\frac{\alpha^{2}}{\pi}\biggr)^{\frac{1}{2}}\exp\Bigl[-\alpha^{2}x^{2}-\frac{p^{2}}{\alpha^{2}\hbar^{2}}\Bigr]\\
	&\qquad\qquad\qquad\qquad\times\int\!dy\;\exp\Bigl[-\frac{\alpha^{2}}{4}\Bigl(y+i\,\frac{2p}{\alpha^{2}\hbar}\Bigr)^{2}\Bigr]\,H_{n}(\alpha (x-\frac{y}{2}))H_{n}(\alpha (x+\frac{y}{2}))\,.\notag
\end{align}
Introduce new variables
\begin{align}
	z&=\frac{\alpha}{2}\Bigl(y+i\,\frac{2p}{\alpha^{2}\hbar}\Bigr)=\frac{\alpha}{2}\,y+i\,\frac{p}{\alpha\hbar}\,,&\beta&=i\,\frac{p}{\alpha\hbar}\,,
\end{align}
and the Wigner function reduces to
\begin{align}
	W_{n}(x,p)&=\frac{(-1)^{n}}{\pi^{\frac{3}{2}}\hbar}\frac{1}{2^{n}n!}e^{-\alpha^{2}x^{2}+\beta^{2}}\int\!dz\;e^{-z^{2}}H_{n}(z-\beta-\alpha x)H_{n}(z-\beta+\alpha x)\,.
\end{align}
With the help of the identity~\cite{GR15}
\begin{equation}
	\int_{-\infty}^{\infty}\!dx\;e^{-x^{2}}H_{m}(x+y)H_{n}(x+z)=2^{n}\pi^{\frac{1}{2}}m!\,z^{n-m}L_{m}^{n-m}(-2yz)\,,
\end{equation}
for $m\leq n$, where $L_{n}^{a}(z)$ is the generalized Laguerre polynomial, we obtain
\begin{equation}
	\int\!dz\;e^{-z^{2}}H_{n}(z-\beta-\alpha x)H_{n}(z-\beta+\alpha x)=2^{n}\pi^{\frac{1}{2}}n!\,L_{n}(2(\alpha^{2}x^{2}-\beta^{2}))\,,
\end{equation}
and then the Wigner function of $n^{\text{th}}$ excited state of the harmonic oscillator becomes
\begin{align}\label{E:gbkseusd}
	W_{n}(x,p)=\frac{(-1)^{n}}{\pi\hbar}\,\exp\Bigl[-\frac{2\mathcal{H}}{\hbar\omega}\Bigr]\,L_{n}(\frac{4\mathcal{H}}{\hbar\omega})\,,
\end{align}
where
\begin{align}
	\alpha^{2}x^{2}-\beta^{2}&=\frac{2\mathcal{H}}{\hbar\omega}\,,&\mathcal{H}&=\frac{p^{2}}{2m}+\frac{m\omega^{2}}{2}x^{2}\,.
\end{align}
The ratio $\mathcal{H}/\omega$ is related to action variable and is an adiabatic invariant in classical mechanics. In quantum physics, the relation
\begin{equation}\label{E:dksdfbsd}
	\mathcal{H}=(n+1/2)\hbar\omega\,,
\end{equation}
roughly defines the boundary between the classically allowed and forbidden regions for the $n^{\text{th}}$ excited state. That is, the phase-space point $(q,p)$ such that $\mathcal{H}>(n+1/2)\hbar\omega$ will fall in the classically forbidden region. The Wigner function $W_{n}$ can be negative when $n\geq1$, in contrast to the probability density $P_{n}$, which in fact is related to the Wigner function by an integral relation
\begin{equation}
	P_{n}(x)=\int\!dp\;W_{n}(x,p)\,.
\end{equation}
Qualitatively speaking, from Fig.~\ref{Fi:harmonOs-2}, the Wigner function of the $n^{\text{th}}$ excited state of the harmonic oscillator will have negative values in the regions that form $n$ among $(2n+1)$ concentric annuli centered at the origin of the phase space. Each annulus has an area roughly order of $\pi\hbar$, so that the total area of negative-value region is of order $n\pi\hbar$. This has a few interesting implications. First in the limit $\hbar\to0$, the areas where the Wigner function takes on negative values have measure zero. Hence essentially the Wigner function becomes non-negative. This seems consistent with the interpretation of identifying the Wigner function as a probability distribution. On the other hand, as $n\gg1$, the total area of the regions the Wigner function takes on negative values increases with $n$, so the Wigner function still keeps the quantum features, and it can never serve as a probability distribution. Hence here we see an example that the large $n$ limit does not always lead to a classical description. In this case, the disparity can be seen from \eqref{E:gbkseusd} that $n$ and $\hbar$ do not appear together as a ratio of the form like $\hbar/n$.

Quantitatively in the limit $n\gg1$, the Wigner function is approximately give by 
\begin{equation}\label{E:dksfhddsf}
	\lim_{n\to\infty}W_{n}(q,p)\simeq\frac{(-1)^{n}}{2\pi^{\frac{3}{2}}\hbar}\biggl(\frac{\hbar\omega}{n\mathcal{H}}\biggr)^{\frac{1}{4}}\biggl[\cos4\sqrt{\frac{n\mathcal{H}}{\hbar\omega}}+\sin4\sqrt{\frac{n\mathcal{H}}{\hbar\omega}}\biggr]+\mathcal{O}(n^{-\frac{3}{4}})\,,
\end{equation}
for a highly excited state, as long as $q$, $p$ are not too close to the boundary defined by \eqref{E:dksdfbsd}, which is related to the turning points of the harmonic potential. Again, it shows that the Winger function remains oscillatory between the positive and negative values within the classically allowed region, even in the $n\gg1$ limit. In comparison, we check the large $\mathcal{H}$ limit, that is, the large energy limit
\begin{equation}
	\lim_{H\to\infty}W_{n}(q,p)\simeq \frac{1}{\pi\hbar n!}\biggl(\frac{4\mathcal{H}}{\hbar\omega}\biggr)^{n}e^{-\frac{2\mathcal{H}}{\hbar\omega}}+\mathcal{O}(\mathcal{H}^{n-1})\,.
\end{equation}
This consistently describes the behavior of the Wigner function in the classically forbidden region. The probability is exponentially suppressed

The semi-classical limit $\hbar\to0$ is rather tricky, and the rigorous treatment of the Wigner function in the semi-classical limit can be found in~\cite{Be77, BM72}. Here we merely discuss a few subtlety in taking the semi-classical limit. Although the total area of the regions where the Wigner function is negative approaches zero in this limit, the transition is rather extreme. From the functional form of \eqref{E:gbkseusd}, we observe that when $\hbar$ is shrunk by a factor $\kappa>1$, the lateral dimension, as seen in Fig.~\ref{Fi:harmonOs-1}-(b), will be squeezed into $1/\sqrt{\kappa}$ of what it was, but the oscillation amplitude is blown up by a factor $\kappa$. For a fixed $n$, the number of oscillations does not change with varying $\hbar$.

\begin{figure}
\centering
    \scalebox{0.45}{\includegraphics{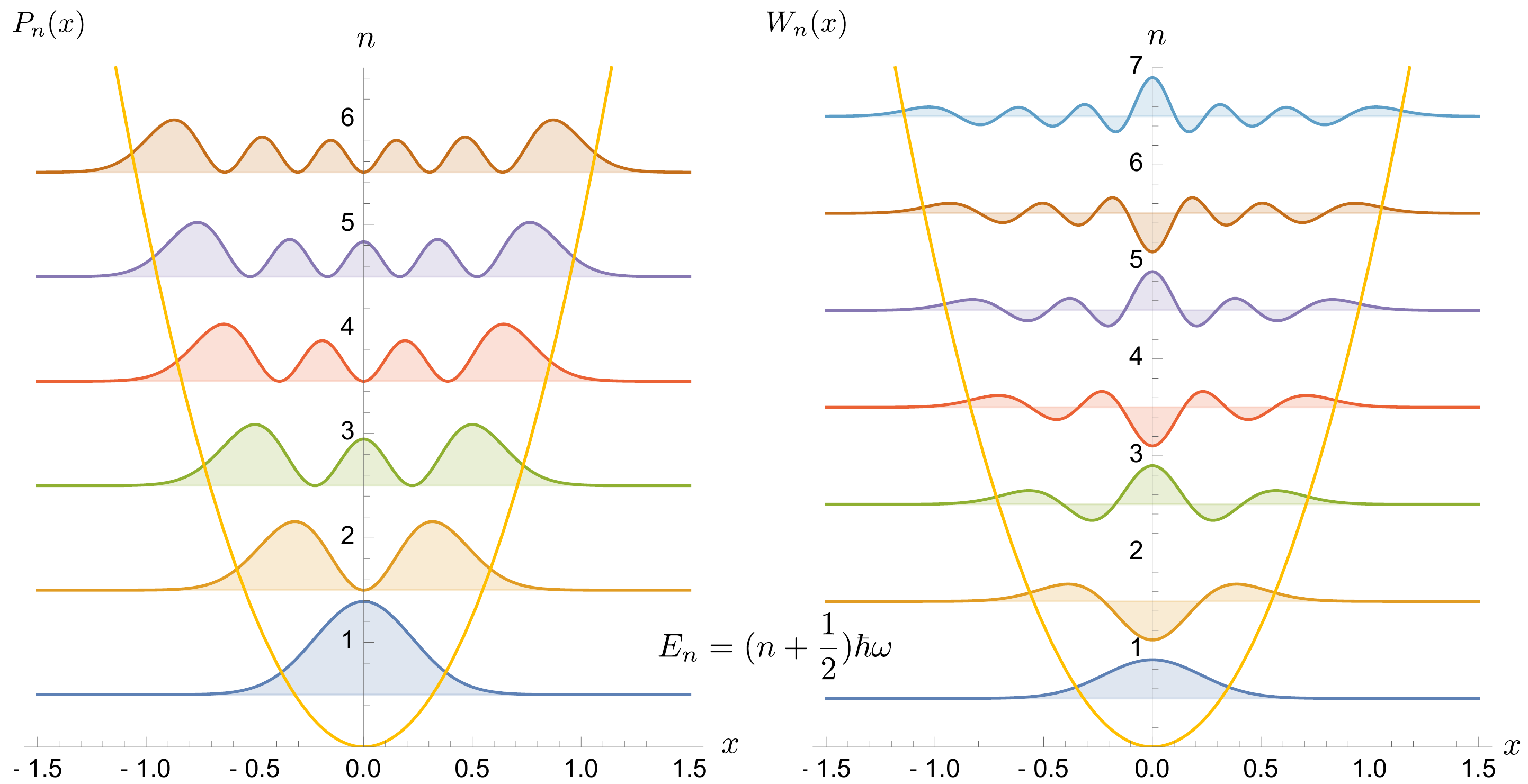}}
    \caption{Comparison of the probability distribution $P_n(x)$ and the Wigner function $W_{n}(x,p=0)$ of a harmonic oscillator from the ground state up to 5th excited state. A few distinguished features of the Wigner function can be easily identified, such as $W_{n}(0)=\frac{(-1)^{n}}{\pi\hbar}$ and the number of the regions where the Wigner function is negative. Both exhibit exponential decay into the classically forbidden regimes.}\label{Fi:harmonOs-2}
\end{figure}

Na\"ively expanding the Wigner function with respect to small $\hbar$ will yield an expression like \eqref{E:dksfhddsf}, which is already positive. Then taking $\hbar\to0$ leads to a limiting form that is proportional to $\delta(x)\delta(p)$, with the sign of the proportionality constant independent of $n$. This is rather perplexing because we know $W_{n}(0)=(-1)^{n}/(\pi\hbar)$. Either the intermediate Taylor expansion \eqref{E:dksfhddsf} or the final $\hbar\to0$ limiting form of the Wigner function do not correctly describe the aforementioned scaling behavior of the Wigner function. Physically, we would expect that if the Wigner function were indeed to give a consistent (semi-)classical description of the quantum harmonic oscillator in phase space when $\hbar\to0$, it would sharply peak at the ellipse, defined by \eqref{E:dksdfbsd}. Clearly the aforementioned scaling behavior does not signal that either. In fact, an implicit feature hints that way. Since the Wigner function is normalized to unity, it means there will be an excess of positive part of the Wigner function over its negative part. The surplus resides on the positive ridge, which is independent of $n$, roughly along the boundary of the classically allowed/forbidden regimes, as seen in Fig.~\ref{Fi:harmonOs-2}. In addition, it has been raised that the more satisfactory way to take the semi-classical limit is to simultaneously require that $n\to\infty$, $\hbar\to0$ but $\mathcal{E}_{n}=(n+\frac{1}{2})\hbar\omega$ should be fixed.  It has been shown~\cite{TZ00} that in such a limit the Wigner function~\eqref{E:gbkseusd} does reduce to
\begin{equation}\label{E:gbddcdds}
	W_{n}(x,p)=\frac{1}{2\pi}\,\delta(\mathcal{H}-E_{n})\,,
\end{equation}
consistent with the classical expectation. {To be thorough, we include the derivation of \eqref{E:gbddcdds} in Appendix~\ref{S:nxkje}.} Here we have used the harmonic oscillator to illustrate the subtleties in and the inequivalence between different approaches of taking the classical limit. Next we will move on to the models that have been used in the context of classicality of cosmological perturbations. In contrast to the harmonic oscillator, they do not have a confining potential, so their canonical coordinate or momentum dispersion tends to grow indefinitely.

\subsection{free particle}\label{S:erikds}
The free particle is used in~\cite{KP98} to highlight features of cosmological perturbation in inflationary spacetime.

Suppose we have a free particle of mass $m$ and it has an initial momentum $p(0)=p_{0}$ at $t=0$, and we further assume that initially its wave function is described by a Gaussian wavepacket such that
\begin{align}
	\langle\hat{x}(0)\rangle&=x_{0}\,,&\langle\hat{p}(0)\rangle&=p_{0}\,,\\
	\langle\Delta\hat{x}^{2}(0)\rangle&=\sigma_{0}^{2}\,,&\langle\Delta\hat{p}^{2}(0)\rangle&=\frac{1}{4\sigma_{0}^{2}}\,,&\frac{1}{2}\langle\bigl\{\Delta\hat{x}(0),\Delta\hat{p}(0)\bigr\}\rangle&=0\,.\label{E:cjmnxde}
\end{align}
The parameter $\sigma_{0}$ denotes the initial width of the state. Then in the Heisenberg picture, By solving the Heisenberg equation
\begin{equation}
	\ddot{\hat{x}}(t)=0\,,
\end{equation}
the position and the momentum operators evolve with time according to
\begin{align}\label{E:kgsdfdsf}
	\hat{x}(t)&=d_{1}(t)\,\hat{x}(0)+\frac{d_{2}(t)}{m}\,\hat{p}(0)\,,&\hat{p}(t)&=m\dot{d}_{1}(t)\,\hat{x}(0)+\dot{d}_{2}(t)\,\hat{p}(0)\,,
\end{align}
where
\begin{align}
	d_{1}(t)&=1\,,&d_{2}(t)&=t\,.
\end{align}
Then from \eqref{E:gksjgs}, we readily find
\begin{align}
	X(t)&=\langle\hat{x}(t)\rangle=x_{0}+\frac{t}{m}\,p_{0}\,,&P(t)&=\langle\hat{p}(t)\rangle=p_{0}\,,\\
	b(t)&=\langle\Delta\hat{x}^{2}(t)\rangle=\sigma_{0}^{2}+\frac{t^{2}}{4m^{2}\sigma_{0}^{2}}\,,&a(t)&=\langle\Delta\hat{p}^{2}(t)\rangle=\frac{1}{4\sigma_{0}^{2}}\,,&c(t)&=\frac{1}{2}\langle\bigl\{\Delta\hat{x}(t),\Delta\hat{p}(t)\bigr\}\rangle=\frac{t}{4m\sigma_{0}^{2}}\,.\label{E:bfhgr}
\end{align}
Together we can explicitly show that the Robertson-Schr\"odinger uncertainty relation is invariant with time
\begin{equation}
	\mathfrak{S}(t)=\mathfrak{S}(0)\,,
\end{equation}
a necessary requirement of a pure quantum state under unitary evolution, where $\mathfrak{S}(t)=ab-c^{2}$,
\begin{align}
	a(t)&=\langle\Delta\hat{p}^{2}(t)\rangle\,,&b(t)&=\langle\Delta\hat{x}^{2}(t)\rangle\,,&c(t)&=\frac{1}{2}\langle\bigl\{\Delta\hat{x}(t),\Delta\hat{p}(t)\bigr\}\rangle\,.
\end{align}
This is of particular importance because it stresses that nothing is lost about the quantumness of the system during the evolution.

The mean position follows the trajectory of the classical free particle and the mean momentum is a constant. They are all consistent with the classical theory. On the other hand, The wavepacket spreads out rapidly in space, which is a very quantum mechanical feature. The position uncertainty grows quadratically indefinitely with time, but the momentum uncertainty remains constant. They, together with the correlation between $\hat{x}$ and $\hat{p}$, ensure that the uncertainty function $\mathfrak{S}(t)$ remains independent of time. It is important to have this correlation involved. Without it, the product $\langle\Delta\hat{x}^{2}(t)\rangle\langle\Delta\hat{p}^{2}(t)\rangle$ grows without bounds.

We also observe that the wavepacket becomes increasingly more squeezed in $\Delta\hat{p}$ with time because the ratio $\langle\Delta\hat{p}^{2}(t)\rangle/\langle\Delta\hat{x}^{2}(t)\rangle$ rapidly diminishes in time even though $\langle\Delta\hat{p}^{2}(t)\rangle$ remains a constant. Strictly speaking, here we have $c\neq0$, so $\Delta\hat{x}$ and $\Delta\hat{p}$ are not orthogonal and they are correlated, so to better describe the deformation of the wavepacket we will find two orthogonal quadratures. They can be determined by the eigenvectors of the covariance matrix for the canonical variables $(\hat{x},\hat{p})$. Following the procedures outlined in \eqref{E:fgkjerd}--\eqref{E:dgsierie} and using the new set of canonical operator pair $(\hat{\mathsf{x}},\hat{\mathsf{p}})$, we find that $c'=0$ always but in the large time limit
\begin{align}\label{E:fgdfrjf}
	b'=\langle\Delta\hat{\mathsf{x}}^{2}\rangle&=\frac{m^{2}\sigma_{0}^{2}}{t^{2}}+\mathcal{O}(\frac{1}{t^{3}})\,,&a'=\langle\Delta\hat{\mathsf{p}}^{2}\rangle&=\frac{t^{2}}{4m^{2}\sigma_{0}^{2}}+\mathcal{O}(t^{0})\,,
\end{align}
while in the short-time limit, we have
\begin{align}
	\langle\Delta\hat{\mathsf{x}}^{2}\rangle&=\sigma_{0}^{2}+\mathcal{O}(t^{2})\,,&\langle\Delta\hat{\mathsf{p}}^{2}\rangle&=\frac{1}{4\sigma_{0}^{2}}+\mathcal{O}(t^{2})\,.
\end{align}
Thus the wavepacket, according to the Wigner function in phase space, is squeezed in the $\hat{\mathsf{x}}$ quadrature but stretched in the $\hat{\mathsf{p}}$ quadrature. It peaks at $(X,P)$, and in particular $X(t)=\langle\hat{x}(t)\rangle=x_{0}+\dfrac{p_{0}}{m}\,t$. On the other hand, since the ellipse defined by these quadratures becomes extremely thin in the $\hat{\mathsf{x}}$ direction but strung out in the $\hat{\mathsf{p}}$ direction. The ellipse seems morphed into a well defined path. However, it does not coincide with the bona fide classical path $C(t)=(x_{0}+\dfrac{p_{0}}{m}\,t,p_{0})$ of a free particle in phase space. Alternatively, since the Wigner function of a free particle can be given by \eqref{E:dgbsjkdhf}, with $a$, $b$, and $c$ in \eqref{E:bfhgr}, at late times we have $b\to\infty$, and we can apply the limit in \eqref{E:ghdisue}. Nonetheless, we note that the path defined by the delta function\footnote{Although in~\cite{KP98}, the authors did not explicitly write the Wigner function into a form proportional to a delta function for the free particle case, their Eq.~(63) and Fig.~1 served the same end. Besides, the Wigner function of the cosmological perturbations in their Eq.~(29) takes the delta-function form. They obtained their Eqs.~(15) and (16)  by keeping only the dominant contributions. According to the analysis in our Sec.~\ref{S:bkgd} these results in~\cite{KP98} are thus problematic.}
\begin{align}\label{E:dgksdjse}
	p-P&=\frac{c}{b}\,\bigl(x-X\bigr)\,,
\end{align}
again in general is not equal to the phase-space path $C(t)$ of the free particle, with an exception of $x=X$ and $p=P$. This special case identically satisfies \eqref{E:dgksdjse}. Finally, according the discussion following \eqref{E:ghdisue}, we learn that the resulting Wigner function is not physical because 1) the pure state is not pure any more after we take the $b\to\infty$ limit, 2) the purity of the state is greater than unity, and 3) the corresponding von Neumann entropy is ill defined. Thus the assumed emergence of a well-defined path in phase space at late times does not lead to a description consistent with the known classical dynamics. What is worse is that the corresponding Wigner function does not even describe a legitimate physical state of a quantum-mechanical system.

The ratio of the square root of the position uncertainty and the mean position gives a measure about the strength of the quantum fluctuations. This is clearly seen from the definition of the position uncertainty. For the free particle, it is given by
\begin{equation}
	\frac{\sqrt{\langle\Delta\hat{x}^{2}(t)\rangle}}{\lvert\langle\hat{x}(t)\rangle\rvert}=\frac{\sqrt{\sigma_{0}^{2}+\frac{t^{2}}{4m^{2}\sigma_{0}^{2}}}}{\lvert x_{0}+\dfrac{t}{m}\,p_{0}\rvert}\,.
\end{equation}
At large times, the ratio reaches to a nonzero constant
\begin{equation}
	\lim_{t\gg1}\frac{\sqrt{\langle\Delta\hat{x}^{2}(t)\rangle}}{\lvert\langle\hat{x}(t)\rangle\rvert}=\frac{1}{2p_{0}\sigma_{0}}+\mathcal{O}(\frac{1}{t})\,.
\end{equation}
so the system still possesses a quantum-fluctuation feature. Since this measure works also for the non-Gaussian state, it cannot be fully accounted for by assuming the Wigner function as a classical probability distribution.

Eq.~\eqref{E:kgsdfdsf} offers an intuitive way to examine the non-commutativity of, say $\hat{x}$, at different times. In the free-particle case, we have
\begin{align}
	\bigl[\hat{x}(t),\hat{x}(t')\bigr]&=\frac{1}{m}\Bigl[d_{1}(t)d_{2}(t')-d_{1}(t')d_{2}(t)\Bigr]\bigl[\hat{x}(0),\hat{p}(0)\bigr]=-i\,\frac{t-t'}{m}\,,\label{E:gbhsejrf}
\end{align}
from \eqref{E:eurbdfhd}. The righthand side is far from being zero, so $\hat{x}$ at different times are strongly non-commuting. From the aforementioned Together with the equal-time commutation relation
\begin{equation}
	\bigl[\hat{x}(t),\hat{p}(t)\bigr]=i\,,
\end{equation}
for all times, we find that the free-particle system remains quantum mechanical throughout the unitary evolution, and the assertion that transition to its classical counterpart at late times cannot be fully justified.

Oftentimes one might resort to the arguments that since $d_{2}(t)\gg d_{1}(t)$ at late times, one might write
\begin{align}\label{E:drejdf}
	\hat{x}(t)\simeq\frac{d_{2}(t)}{m}\,\hat{p}(0)
\end{align}
at late times, such that one would conclude
\begin{equation}
	\bigl[\hat{x}(t),\hat{x}(t')\bigr]\simeq\bigl[\frac{d_{2}(t)}{m}\,\hat{p}(0),\frac{d_{2}(t')}{m}\,\hat{p}(0)\bigr]=0\,,
\end{equation}
and claim that the operators becomes commuting. A more careful analysis based on \eqref{E:gbhsejrf} plainly shows that the approximation used in \eqref{E:drejdf} is misleading and the subsequent conclusion is then illusory.

Next we will examine the quantum inverted oscillator used in~\cite{GuthPi85}, which in some sense mimics the runaway behavior of the cosmological perturbations in the de Sitter space.

\subsection{inverted linear oscillator}\label{S:erikds1}
The inverted oscillator's potential has an opposite sign to the harmonic potential, so the Heisenberg equation takes the form
\begin{equation}\label{E:epoird}
	\ddot{\hat{x}}(t)-\omega^{2}\hat{x}(t)=0\,.
\end{equation}
Strictly speaking, its motion is not oscillatory, so the parameter $\omega>0$ does not bear the meaning of oscillation frequency. In general, the classical inverted oscillator has unstable, runaway dynamics, except for the occasion that the initial conditions satisfy $\dot{x}(0)+\omega\,x(0)=0$. An example is the case when the system initially rests at the top of the potential. The quantum inverted oscillator is susceptible to its own quantum fluctuations, so its dynamics is more prone to run away.

The fundamental solutions to \eqref{E:epoird} are
\begin{align}\label{E:fgsjh}
	d_{1}(t)&=\cosh\omega t\,,&d_{2}(t)&=\frac{1}{\omega}\,\sinh\omega t\,.
\end{align}
Suppose the initial conditions are given by
\begin{align}
	\langle\hat{x}(0)\rangle&=0\,,&\langle\hat{p}(0)\rangle&=0\,,&\langle\hat{x}^{2}(0)\rangle&=\frac{1}{4\beta}\,,&\langle\hat{p}^{2}(0)\rangle&=\beta\,,&\frac{1}{2}\langle\bigl\{\hat{x}(0),\hat{p}(0)\bigr\}\rangle&=0\,.
\end{align}
Then from \eqref{E:gksjgs}, we immediately have
\begin{align}
	b&=\langle\Delta\hat{x}^{2}(t)\rangle=\frac{\cosh2\omega t+\cos2\theta}{8\beta\cos^{2}\theta}\,,&&\text{and as $t\gg\omega^{-1}$}&\langle\Delta\hat{x}^{2}(t)\rangle&=\frac{1}{16\beta\cos^{2}\theta}\,e^{2\omega t}\,,\label{E:djgvejh1}\\
	a&=\langle\Delta\hat{p}^{2}(t)\rangle=\frac{\beta}{2}\frac{\cosh2\omega t-\cos2\theta}{\sin^{2}\theta}\,,&&\text{and as $t\gg\omega^{-1}$}&\langle\Delta\hat{p}^{2}(t)\rangle&=\frac{\beta}{4\sin^{2}\theta}\,e^{2\omega t}\,,
\intertext{and their cross correlation}
	c&=\frac{1}{2}\langle\bigl\{\Delta\hat{x}(t),\Delta p(t)\bigr\}\rangle=\frac{\sinh2\omega t}{4\sin\theta\cos\theta}\,,&&\text{and as $t\gg\omega^{-1}$}&\frac{1}{2}\langle\bigl\{\Delta\hat{x}(t),\Delta p(t)\bigr\}\rangle&=\frac{1}{8\sin\theta\cos\theta}\,e^{2\omega t}\,,\label{E:djgvejh3}
\end{align}
with the mean values given by
\begin{align}
	\langle\hat{x}(t)\rangle&=0\,,&\langle\hat{p}(t)\rangle&=0\,.
\end{align}
The parameters $\beta$, $\theta$ are chosen such that~\cite{GuthPi85}
\begin{equation}
	\beta=\frac{m\omega}{2}\,\tan\theta>0\,.
\end{equation}
The former is associated with the width of the initial wavefunction, and comparing with \eqref{E:cjmnxde}, we may identify $\beta=1/(4\sigma_{0}^{2})$. The mean energy of the system is conserved and is given by
\begin{equation}
	\langle\hat{H}(t)\rangle=\frac{\langle\hat{p}^{2}(t)\rangle}{2m}-\frac{m\omega^{2}}{2}\,\langle\hat{x}^{2}(t)\rangle=-\frac{\omega}{2}\,\cot2\theta\,,
\end{equation}
in which the kinetic energy $E_{k}$ and the potential energy $E_{p}$ are
\begin{align}
	E_{k}(t)&=\frac{\langle\hat{p}^{2}(t)\rangle}{2m}=\frac{\cosh2\omega t-\cos\theta}{8\sin\theta\cos\theta}\,\omega\,,&E_{p}(t)&=-\frac{m\omega^{2}}{2}\,\langle\hat{x}^{2}(t)\rangle=-\frac{\cosh2\omega t+\cos\theta}{8\sin\theta\cos\theta}\,\omega\,.
\end{align}
Since the potential is unbounded below, we expect the kinetic energy of the inverted oscillator will also increase without bound.

Thus although averagely speaking, the mean position of the quantum inverted oscillator remains at the top of the potential, its position dispersion spreads exponentially fast. It means when we try to measure the coordinate operator $\hat{x}$, it is more than often that we will obtain a nonzero value, and with increasing time, the typical measured value grows indefinitely, {rolling} down under  either side of the inverted potential. Therefore in the coordinate representation, the wavefunction is widely spread about the mean trajectory. The same conclusion applies to the measurement of the canonical momentum operator $\hat{p}$. These also hold true even when the initial conditions are such that the system rolls downs that potential. From \eqref{E:fgeusdfg} and \eqref{E:fgsjh}, we see that the mean position will increase exponentially, but the position dispersion $\sqrt{\langle\Delta\hat{x}^{2}(t)\rangle}$ grows equally fast. Even the system has such a runaway behavior, its uncertainty function $\mathfrak{S}$ remains a constant value
\begin{equation}
	\mathfrak{S}(t)=\langle\Delta\hat{x}^{2}(t)\rangle\langle\Delta\hat{p}^{2}(t)\rangle-\biggl[\frac{1}{2}\langle\bigl\{\Delta\hat{x}(t),\Delta p(t)\bigr\}\rangle\biggr]^{2}=\frac{1}{4}\,.
\end{equation}
Again we emphasize that this reveals that the quantum nature of the system never loses and decoherence does not happen, even though the system may seem to show classical behavior in some partial measures. In contrast, if one uses the approximated forms of the covariance matrix elements in \eqref{E:djgvejh1}--\eqref{E:djgvejh3}, one will obtain a zero value for the uncertainty function, which violates unitarity. It would be misleading if one uses this as the criterion to claim the emergence of classicality due to a loss of quantum coherence. This example also tells that one should be mindful of the contributions of the subdominant contributions in the covariant matrix elements.

The Wigner function takes the standard form for the Gaussian state we used,
\begin{equation}
	\mathcal{W}(x,p;t)=\frac{1}{\pi}\,\exp\Bigl[-2ax^{2}+4c\,xp-2bp^{2}\Bigr]\,,
\end{equation}
with $ab-c^{2}=1/4$. However, in contrast to the free-particle case, here the momentum uncertainty also increases indefinitely with time. Thus it is interesting to examine the behavior of the Wigner function to see whether it will define a highly squeezed ellipse at late times. Discussions in the previous sections indicate that the ratio of the two orthogonal quadratures can provide the information about the extent of squeezing of the ellipse during the course of evolution. From \eqref{E:fgkjerd}, we find the ratio given by
\begin{equation}
	\lim_{t\to\infty}\frac{\lambda_{-}}{\lambda_{+}}=\frac{16\beta^{2}\sin^{4}2\theta}{[4\beta^{2}+1+(4\beta^{2}-1)\cos2\theta]^{2}}\,e^{-4\omega t}+\cdots\,,
\end{equation}
at late times. It falls to zero extremely fast, so we end up with a highly stretched and highly squeezed ellipse. As a reminder, even the ratio takes on such an extreme value, the product $\lambda_{+}\lambda_{-}$ remains $1/4$, a rephrasing of the Robertson-Schr\"odinger uncertainty relation for the orthogonal quadratures. Since from \eqref{E:ghdisue}, if we take the limit $b\to\infty$ too literally, the Wigner function contains a delta function that defined a path in phase space\footnote{In~\cite{GuthPi85}, only the dominant contribution is kept, so the  derived Wigner function of a quantum inverted oscillator in their Eq.~(2.13) contains a delta function. It does not describe a physical state,  according to~\cite{EF98} and the reasoning in our Sec.~\ref{S:bkgd}.},
\begin{equation}
	p-\frac{c}{b}\,x=0\,.
\end{equation}
Putting the values of $b$ and $c$ for the inverted oscillator in \eqref{E:djgvejh1} and \eqref{E:djgvejh3}, we obtain
\begin{align}\label{E:rhfd}
	p&=m\omega\,\frac{\sinh2\omega t}{\cosh2\omega t+\cos\theta}\,x\simeq m\omega\,x\,,&&\text{as}&t\gg\omega^{-1}\,.
\end{align}
This, in the current case when $\langle\hat{x}\rangle=0$, $\langle\hat{p}\rangle=0$, clearly does not match the classical counterpart even in the limit $t\to\infty$. One may find this example too atypical, so let us consider the initial conditions of the inverted oscillator such that its classical counterpart does roll down that potential. The mean position and momentum are given by
\begin{align}\label{E:gbsjcxc}
	X(t)&=\cosh\omega t\,X(0)+\frac{1}{m\omega}\,\sinh\omega t\,P(0)\,,&P(t)&=m\omega\,\sinh\omega t\,X(0)+\cosh\omega t\,P(0)\,.
\end{align}
Eq.~\eqref{E:rhfd} becomes
\begin{align}
	p=\frac{c}{b}\,x+\Bigl[P(t)-\frac{c}{b}\,X(t)\Bigr]\,.
\end{align}
This does not resemble the classical path $(X(t),P(t))$ in phase space. Following our discussions in the free-particle case in treating a highly squeezed ellipse, one should not fall into the trap of misinterpreting this as the emergence of classicality of the quantum system. Here we would like to stress again that no matter how one deforms the quadrature ellipse in phase space by squeezing and stretching, it is always a two-dimensional geometric object, having an invariant area $\pi\hbar/4$, a dictum of the quantum uncertainty principle and unitary evolution of a closed system. Nonetheless when one jumps to reducing the ellipse to a line in phase space, the areas go to zero, and unitarity is violated.

Here it is also interesting to note that suppose we have two ``density matrix elements'', and they differ by
\begin{equation}
	\varrho(x,x';t)=\rho(x,x';t)\,\exp\Bigl[\gamma\bigl(x-x'\bigr)^{2}\Bigr]
\end{equation}
for some real number $\gamma$ and have the property $\varrho(x,x;t)=\rho(x,x;t)$. Then these two density matrix elements will give the same probability distribution and satisfy the same normalization condition, but they may not all be physical. They may not be guaranteed to be semi-positive definite.

Finally we examine the commutator of the position operator at different times
\begin{align}
	\bigl[\hat{x}(t),\hat{x}(t')\bigr]=\frac{1}{m}\bigl[d_{1}(t)d_{2}(t')-d_{1}(t')d_{2}(t)\bigr]\,\bigl[\hat{x}(0),\hat{p}(0)\bigr]=-\frac{i}{m\omega}\,\sinh\omega(t-t')\,.
\end{align}
The commutator is far from being zero, so they do not commute. For the motion of the inverted oscillator having the mean position and momentum, given by \eqref{E:gbsjcxc}, the effects of quantum fluctuation effect are not negligible, because the ratio
\begin{equation}
	\lim_{t\to\infty}\frac{\sqrt{\langle\Delta\hat{x}^{2}(t)\rangle}}{\langle\hat{x}(t)\rangle}=\frac{m\omega}{p_{0}}\sqrt{\frac{1}{4\beta}+\frac{\beta}{m^{2}\omega^{2}}}\,.
\end{equation}
approaches a constant of order unity, as $t\to\infty$. Therefore as in the free particle case, we conclude that the quantum inverted oscillator remains quantal throughout its evolution.

To summarize the salient features in our model studies discussed in this section 
we refer the reader  back to Sec.~\ref{S:ribdfgd}, for an itemized list of pivotal findings.

{In the next section we shall discuss the evolution of a quantum field in an inflationary universe. After a long duration of inflation  the quantum state has undergone a high degree of squeezing. This is the reason why some authors 
felt justified enough to adopt the leading-order approximation, and, without providing any convincing explanation, ignored the sub-leading terms.  With the help of the  simpler quantum mechanical models  studied in this and the previous section,  we have pin-pointed where the pitfalls are with regard to the classicalization issue   
where the fallacy of the prior claims resides. In the same vein, for the case of inflaton studied in the next section
we shall show that viewed as a closed system, quantum cosmological perturbations do not decohere. In addition, the existence of quantum entanglement lends support to our thesis that closed  quantum systems do not  turn classical just because they are badly squeezed.}

\section{inflaton field}\label{S:ebjfkd}
Cosmological perturbations in an inflationary universe, in an appropriate gauge, follows the linear perturbations of the inflaton field~\cite{IntEnt}, so we will just follow the unitary evolution of the perturbations of the inflaton field.

The perturbation of the inflaton field can be described by a minimally coupled scalar field $\phi$ in spatially flat de Sitter space, whose line element in the conformal time frame is given by
\begin{equation}
	ds^{2}=a^{2}(\eta)\,\bigl(-d\eta^{2}+dx_{i}^{2}\bigr)\,,
\end{equation}
Here $a(\eta)$ is the scale factor that depends only on the conformal time $\eta$. Thus the Lagrangian takes the form
\begin{align}
	L&=-\frac{1}{2}\int\!d^{3}x\;a^{2}\Bigl[-\bigl(\partial_{\eta}\phi\bigr)^{2}+\bigl(\partial_{i}\phi\bigr)^{2}\Bigr]=\frac{a^{2}(\eta)}{2}\sum_{\bm{k}}\Bigl[\phi'_{\bm{k}}(\eta)\phi'^{*}_{\bm{k}}(\eta)-\bm{k}^{2}\phi^{\vphantom{*}}_{\bm{k}}(\eta)\phi^{*}_{\bm{k}}(\eta)\Bigr]\,,
\end{align}
where we decompose the field into its modes, and the temporal dependence is included in $\phi_{\bm{k}}(\eta)$
\begin{equation}
	\phi(\bm{x},\eta)=\sum_{\bm{k}}\phi_{\bm{k}}(\eta)\,e^{+i\bm{k}\cdot\bm{x}}\,.
\end{equation}
It is convenient to make a change of variable $\chi_{\bm{k}}(\eta)=a(\eta)\,\phi_{\bm{k}}(\eta)$, and write the Lagrangian as\footnote{Or in the coordinate representation, we have
\begin{equation*}
	L=\frac{1}{2}\int\!d^{3}x\;\Bigl\{\chi'^{2}(\bm{x},\eta)-\bigl[\partial_{i}\chi(\bm{x},\eta)\bigr]^{2}+\frac{a''}{a^{2}}\,\chi^{2}(\bm{x},\eta)-\frac{d}{d\eta}\Bigl[\frac{a'(\eta)}{a(\eta)}\,\chi(\bm{x},\eta)\Bigr]^{2}\Bigr\}\,,
\end{equation*}
and
\begin{equation*}
	p(\bm{x},\eta)=\chi'(\bm{x},\eta)\,.
\end{equation*}
}
\begin{align}
	L&=\frac{1}{2}\sum_{\bm{k}}\Bigl\{\chi'_{\bm{k}}\chi'^{*}_{\bm{k}}-\Bigl[\bm{k}^{2}-\frac{a''}{a}\Bigr]\,\chi^{\vphantom{*}}_{\bm{k}}\chi^{*}_{\bm{k}}-\frac{d}{d\eta}\Bigl[\frac{a'}{a}\,\chi^{\vphantom{*}}_{\bm{k}}\chi^{*}_{\bm{k}}\Bigr]\Bigr\}\,.
\end{align}
The last term is a total time derivative, so its contribution can be discarded. Thus the canonical momentum conjugated to $\chi_{\bm{k}}$ is given by
\begin{equation}
	p_{\bm{k}}=\frac{\partial L}{\partial \chi'_{\bm{k}}}=\chi'^{*}_{\bm{k}}\,,
\end{equation}
and then the equation of motion is
\begin{equation}\label{E:erkfhbd}
	\chi''_{\bm{k}}+\Bigl[\bm{k}^{2}-\frac{a''}{a}\Bigr]\,\chi^{\vphantom{*}}_{\bm{k}}=0\,.
\end{equation}
Each mode essentially behaves like a linear parametric oscillators, so its evolution remains Gaussian if the initial state of such a system is a Gaussian. If the initial state is a pure state like the vacuum state, then it will evolve to a two-mode squeezed (vacuum) state~\cite{IntEnt}. Creation of particle pairs  will accompany such a parametric evolution.

If $a(\eta)=-(H\eta)^{-1}$ with $-\infty<\eta<0^{-}$, then the mode function $u_{k}(\eta)$ is given by
\begin{equation}
	u_{k}(\eta)=\frac{1}{\sqrt{2k}}\,\Bigl(1-\frac{i}{k\eta}\Bigr)\,e^{-ik\eta}\,,
\end{equation}
normalized by the Wronksian condition $u^{\vphantom{*}}_{k}u^{'*}_{k}-u'^{\vphantom{*}}_{k}u^{*}_{k}=i$.  {After we promote the canonical variables to operators, they can be expanded by the mode function $u_{k}(\eta)$,}
\begin{align}\label{E:kdgbddf}
	\hat{\chi}_{\bm{k}}(\eta)&=\hat{a}^{\vphantom{\dagger}}_{+\bm{k}}\,u^{\vphantom{*}}_{k}(\eta)+\hat{a}^{\dagger}_{-\bm{k}}\,u^{*}_{k}(\eta)\,,&\hat{p}_{\bm{k}}(\eta)&=\hat{a}^{\vphantom{\dagger}}_{-\bm{k}}\,u'^{\vphantom{*}}_{k}(\eta)+\hat{a}^{\dagger}_{+\bm{k}}\,u'^{*}_{k}(\eta)\,,
\end{align}
and $\hat{\chi}_{-\bm{k}}^{\vphantom{\dagger}}=\hat{\chi}_{+\bm{k}}^{\dagger}$. In the limit $\eta\to-\infty$, the mode function corresponds to the positive-frequency mode
\begin{align}
	\lim_{\eta\to-\infty}u_{k}(\eta)&=\frac{1}{\sqrt{2k}}\,e^{-ik\eta}\,,&\lim_{\eta\to-\infty}u'_{k}(\eta)&=-i\sqrt{\frac{k}{2}}\,e^{-ik\eta}\,.
\end{align}
However, we may consider a more general case that {the inflation starts} at $\eta=\eta_{0}<0$. We then require that $u_{k}(\eta_{0})=\dfrac{1}{\sqrt{2k}}$ and $u'_{k}(\eta_{0})=-i\sqrt{\dfrac{k}{2}}$. Thus the positive-frequency mode function takes a rather complicated form
\begin{equation}\label{E:fjbdkfs}
	u_{k}(\eta)=\frac{1}{\sqrt{2k}}\,e^{-ik(\eta-\eta_{0})}\biggl\{i\,\frac{(1+ik\eta)[1+(1-i)k\eta_{0}][1-(1+i)k\eta_{0}]}{2k^{3}\eta\eta_{0}^{2}}-i\,\frac{1-ik\eta}{2k^{3}\eta\eta_{0}^{2}}\,e^{+i2k(\eta-\eta_{0})}\biggr\}\,,
\end{equation}
and the fundamental solutions are given by
\begin{align}
	d_{k}^{(1)}(\eta)&=\frac{k[\eta_{0}-\eta(1-k^{2}\eta_{0}^{2})]\cos k(\eta-\eta_{0})+(1+k^{2}\eta\eta_{0})\sin k(\eta-\eta_{0})}{k^{3}\eta\eta_{0}}\,,\\
	d_{k}^{(2)}(\eta)&=\frac{-k(\eta-\eta_{0})\cos k(\eta-\eta_{0})+(1+k^{2}\eta\eta_{0})\sin k(\eta-\eta_{0})}{k^{3}\eta\eta_{0}}\,,
\end{align}
with
\begin{align}
	d_{k}^{(1)}(\eta_{0})&=1\,,&d'^{(1)}_{k}(\eta_{0})&=0\,,&d_{k}^{(2)}(\eta_{0})&=0\,,&d'^{(2)}_{k}(\eta_{0})&=1\,,
\end{align}
Then the general evolution of $\hat{\chi}_{\bm{k}}(\eta)$ can be constructed in the same way as outlined in Sec.~\ref{S:bkgd} by
\begin{align}
	\hat{\chi}_{\bm{k}}(\eta)&=d_{k}^{(1)}(\eta)\,\hat{\chi}_{\bm{k}}(\eta_{0})+d_{k}^{(2)}(\eta)\,\hat{\chi}'_{\bm{k}}(\eta_{0})=d_{k}^{(1)}(\eta)\,\hat{\chi}_{+\bm{k}}(\eta_{0})+d_{k}^{(2)}(\eta)\,\hat{p}_{-\bm{k}}(\eta_{0})\,,\label{E:fhkfhgk1}\\
	\hat{\chi}'_{\bm{k}}(\eta)&=d'^{(1)}_{k}(\eta)\,\hat{\chi}_{\bm{k}}(\eta_{0})+d'^{(2)}_{k}(\eta)\,\hat{\chi}'_{\bm{k}}(\eta_{0})\,,
\end{align}
such that
\begin{equation}\label{E:fhkfhgk3}
	\hat{p}_{\bm{k}}(\eta)=\hat{\chi}'^{\dagger}_{\bm{k}}(\eta)=d^{(1)}_{k}{}'(\eta)\,\hat{\chi}_{-\bm{k}}(\eta_{0})+d^{(2)}_{k}{}'(\eta)\,\hat{p}_{+\bm{k}}(\eta_{0})\,.
\end{equation}
At the initial time the mode operator can be expressed in terms of the creation and annihilation operators $(\hat{a}^{\dagger}_{\pm\bm{k}},\hat{a}^{\vphantom{\dagger}}_{\pm\bm{k}})$ of the $\pm\bm{k}$ modes,
\begin{align}\label{E:gustrdf}
	\hat{\chi}_{\bm{k}}(\eta_{0})&=\frac{1}{\sqrt{2k}}\bigl(\hat{a}^{\dagger}_{-\bm{k}}+\hat{a}^{\vphantom{\dagger}}_{+\bm{k}}\bigr)\,,&\hat{\chi}'_{\bm{k}}(\eta_{0})&=i\sqrt{\frac{k}{2}}\bigl(\hat{a}^{\dagger}_{-\bm{k}}-\hat{a}^{\vphantom{\dagger}}_{+\bm{k}}\bigr)\,.
\end{align}
Then we can express $\hat{\chi}_{\bm{k}}$ at any time in terms of the creation and annihilation operators at the initial times,
\begin{equation}\label{E:dgbksdere}
	\hat{\chi}_{\bm{k}}(\eta)=\frac{1}{\sqrt{2k}}\Bigl[d_{k}^{(1)}(\eta)-ik\,d_{k}^{(2)}(\eta)\Bigr]\hat{a}^{\vphantom{\dagger}}_{+\bm{k}}+\frac{1}{\sqrt{2k}}\Bigl[d_{k}^{(1)}(\eta)+ik\,d_{k}^{(2)}(\eta)\Bigr]\hat{a}^{\dagger}_{-\bm{k}}\,.
\end{equation}
Compared with \eqref{E:kdgbddf} we arrive at a convenient relation 
\begin{equation}\label{E:gbseud}
	u_{k}(\eta)=\frac{1}{\sqrt{2k}}\Bigl[d_{k}^{(1)}(\eta)-ik\,d_{k}^{(2)}(\eta)\Bigr]\,,
\end{equation}
between the mode function $u_{k}(\eta)$ and the fundamental solutions $d_{k}^{(1,2)}(\eta)$ to the differential equation \eqref{E:erkfhbd}.

\subsection{Canonical variables remain noncommutating}

From \eqref{E:fhkfhgk1}, we may compute the commutators of $\hat{\chi}_{\bm{k}}$ at different times\footnote{In the context of a parametrically driven quantum system,   \eqref{E:fgkdjf}   is used for the discussion about the non-commutativity of an operator at different times, not \eqref{E:keijrs}, which is usually used in the non-parametrically driven cases, such as the free particle and the inverted oscillator.}
\begin{align}\label{E:keijrs}
	\bigl[\hat{\chi}_{\bm{k}}(\eta),\hat{\chi}_{\bm{k}}(\eta')\bigr]&=0\,,
	\intertext{but}
	\bigl[\hat{\chi}_{\bm{k}}(\eta),\hat{\chi}^{\dagger}_{\bm{k}}(\eta')\bigr]&=i\,\frac{k(\eta-\eta')\,\cos k(\eta-\eta')+(1+k^{2}\eta\eta')\sin k(\eta-\eta')}{k^{3}\eta\eta'}\,,\label{E:fgkdjf}
\end{align}
independent of $\eta_{0}$, so the same expression even if $\eta_{0}\to-\infty$. The commutator Eq.~\eqref{E:fgkdjf} for the superhorizon mode $k\eta\ll1$, $k\eta'\ll1$ in general does not vanish. It is approximately given by
\begin{equation}
	\bigl[\hat{\chi}_{\bm{k}}(\eta),\hat{\chi}^{\dagger}_{\bm{k}}(\eta')\bigr]=-i\,\frac{\eta^{3}-\eta'^{3}}{3\eta\eta'}\,,
\end{equation}
and will approach zero only if additional limits $\eta\to0^{-}$ and $\eta'\to0^{-}$ are imposed. Thus in general these two operators do not commute.

Notice that if one opts to keep only the growing part contribution of $d_{k}^{(i)}(\eta)$ for the superhorizon modes, then $[\hat{\chi}_{\bm{k}}(\eta),\hat{\chi}^{\dagger}_{\bm{k}}(\eta')]=0$ without additional requirement $\eta$ and $\eta'\to0^{-}$. This is where some prior authors wrongly conclude that these two operators commute for superhorizon modes at any time.

Different from the previous examples, given a fixed $k$, we can always find, in the inflaton field case, a sufficiently small $\eta$ and $\eta'$ to make the commutator $[\hat{\chi}_{\bm{k}}(\eta),\hat{\chi}^{\dagger}_{\bm{k}}(\eta')]$ as close to zero as possible. This is not a typical scenario. On the other hand, the equal-time commutation relation between the canonical variables is always nonzero, given by
\begin{align}\label{E:gbsjkhvd}
	\bigl[\hat{\chi}_{\bm{k}}(\eta),\hat{p}_{\bm{k}}(\eta)\bigr]=\Bigl[d_{k}^{(1)}(\eta)d_{k}^{(2)}{}'(\eta)-d_{k}^{(1)}{}'(\eta)d_{k}^{(2)}(\eta)\Bigr]\,\bigl[\hat{\chi}_{\bm{k}}(\eta_{0}),\hat{p}_{\bm{k}}(\eta_{0})\bigr]=i\,.
\end{align}
A spurious argument is also often used for the equal-time commutation relation in this context: For a finite $\eta_{0}$, if one keeps only the dominant contributions of $d_{k}^{(i)}(\eta)$ for the superhorizon modes
\begin{align}
	d_{k}^{(1)}(\eta)&\simeq\frac{1}{k\eta}\Bigl[\frac{\cos k\eta_{0}}{k\eta_{0}}+\sin k\eta_{0}-\frac{\sin k\eta_{0}}{k^{2}\eta_{0}^{2}}\Bigr]\,,&d_{k}^{(2)}(\eta)&\simeq\frac{1}{k\eta}\Bigl[\frac{\cos k\eta_{0}}{k}-\frac{\sin k\eta_{0}}{k^{2}\eta_{0}}\Bigr]\,,
\end{align}
then one will obtain
\begin{equation}
	d_{k}^{(1)}(\eta)d_{k}^{(2)}{}'(\eta)-d_{k}^{(1)}{}'(\eta)d_{k}^{(2)}(\eta)=0\,,
\end{equation}
because both $d_{k}^{(i)}(\eta)$ have the same {form of $\eta$} dependence. Thus one again incorrectly concludes that
\begin{equation}\label{E:fgdedf}
	\bigl[\hat{\chi}_{\bm{k}}(\eta),\hat{p}_{\bm{k}}(\eta)\bigr]\simeq0
\end{equation}
and wrongly proclaim that the quantum-to-classical transition has implicitly occurred. Eq.~\eqref{E:fgdedf} is clearly in contradiction with \eqref{E:gbsjkhvd}. {In addition, all the issues concerning the Wigner function\footnote{For example, in~\cite{PS96}, the Wigner funciton of the cosmological perturbations in its Eq.~(46) is written into a delta function form, and when only the leading term {of what is resulting in} its Eq.~(51) is kept, the conclusion of `decoherence without decoherence' is conveniently yet haphazardly drawn.} discussed in Sec.~\ref{S:dkhejre} apply here. The extreme squeezing of the state, caused by the exponential expansion of the background spacetime, offers a strong temptation for authors of this  persuasion to dispense with the subleading terms, and to draw physical conclusions based only on the dominant contributions. Thus,} the salient lesson {we have learned so far} is that \textit{the subdominant contributions may not be wantonly discarded and must be treated carefully}. Then we will correctly find that the non-commutativity between the canonical operators are intact and robust.

\subsection{Particle Creation: Numbers and Coherence}

The inflaton field perturbation has a distinguished quantum feature that is absent in the examples in Sec.~\ref{S:dkhejre}. The particles in the $\pm\bm{k}$ modes are created in pair over the parametric evolution driven by the expanding spacetime. These particles are not created incoherently. In fact they are entangled. Even though particles are copiously produced and entangled, the state of the perturbation remains pure without entropy production~\cite{LCH10, IntEnt}.

Let us look into this aspect in more details. In the Heisenberg picture, the time evolution of the linear field operator can be expressed as a mapping of the operators in terms of the squeezed transformation
\begin{align}
	&\hat{a}_{\bm{k}}&&\mapsto &S_{2}^{\dagger}(\zeta_{\bm{k}})\,\hat{a}_{\bm{k}}\,S_{2}^{\vphantom{\dagger}}(\zeta_{\bm{k}})&=\cosh\eta_{\bm{k}}\,\hat{a}_{+\bm{k}}^{\vphantom{\dagger}}-e^{+i\theta_{\bm{k}}}\sinh\eta_{\bm{k}}\,\hat{a}_{-\bm{k}}^{\dagger}\,,
\end{align}
with the two-mode squeeze operator
\begin{equation}
	S_{2}^{\vphantom{\dagger}}(\zeta_{\bm{k}})=\exp\Bigl[\zeta^{*}_{\bm{k}}\hat{a}_{+\bm{k}}^{\vphantom{\dagger}}\hat{a}_{-\bm{k}}^{\vphantom{\dagger}}-\zeta^{\vphantom{*}}_{\bm{k}}\hat{a}_{+\bm{k}}^{\dagger}\hat{a}_{-\bm{k}}^{\dagger}\Bigr]\,,
\end{equation}
and the rotation transformation
\begin{align}
	&\hat{a}_{\bm{k}}&&\mapsto &R^{\dagger}(\psi_{\bm{k}})\,\hat{a}_{\bm{k}}\,R^{\vphantom{\dagger}}(\psi_{\bm{k}})&=\hat{a}_{\bm{k}}^{\vphantom{\dagger}}\,e^{-i\psi_{\bm{k}}}\,,
\end{align}
with the rotation operator
\begin{equation}
	R^{\vphantom{\dagger}}(\psi_{\bm{k}})=\exp\Bigl[-i\,\psi_{\bm{k}}\Bigl(\hat{a}_{\bm{k}}^{\dagger}\hat{a}_{\bm{k}}^{\vphantom{\dagger}}+\frac{1}{2}\Bigr)\Bigr]\,.
\end{equation}
The squeeze parameter $\zeta_{\bm{k}}=\eta_{\bm{k}}\,e^{i\theta_{\bm{k}}}$ and the rotation angle $\psi_{\bm{k}}$ are time-dependent functions, reflecting the time evolution of the operator $\hat{a}_{\bm{k}}$ in this case. At a later time, the operator $\hat{a}_{\bm{k}}$ at the initial time is formally mapped to
 \begin{equation}\label{E:dgbsedf}
	\hat{b}_{+\bm{k}}=R^{\dagger}(\psi_{\bm{k}})S_{2}^{\dagger}(\zeta_{\bm{k}})\,\hat{a}_{\bm{k}}\,S_{2}^{\vphantom{\dagger}}(\zeta_{\bm{k}})R^{\vphantom{\dagger}}(\psi_{\bm{k}})=\alpha^{\vphantom{*}}_{\bm{k}}\,\hat{a}^{\vphantom{\dagger}}_{+\bm{k}}+\beta^{*}_{\bm{k}}\,\hat{a}^{\dagger}_{-\bm{k}}\,,
\end{equation}
with\begin{align}
	\alpha^{\vphantom{*}}_{\bm{k}}&=e^{-i\psi_{\bm{k}}}\cosh\eta_{\bm{k}}\,,&\beta^{*}_{\bm{k}}&=-e^{-i\psi_{\bm{k}}}e^{+i\theta_{\bm{k}}}\sinh\eta_{\bm{k}}\,.
\end{align}
The Bogoliubov coefficients $\alpha^{\vphantom{*}}_{\bm{k}}$, $\beta^{\vphantom{*}}_{\bm{k}}$ obey the Wronskian condition
\begin{align}
	\lvert\alpha^{\vphantom{*}}_{\bm{k}}\rvert^{2}-\lvert\beta^{\vphantom{*}}_{\bm{k}}\rvert^{2}=1\,.
\end{align}
Two useful combinations of the Bogoliubov coefficients are
\begin{align}
	\lvert\beta_{\bm{k}}\rvert^{2}&=\sinh^{2}\eta_{-\bm{k}}=\sinh^{2}\eta_{\bm{k}}\,,\label{E:fkgjhsd1}\\
	\alpha^{\vphantom{*}}_{\bm{k}}\beta_{\bm{k}}^{*}&=-e^{i(\psi_{-\bm{k}}-\psi_{+\bm{k}})}e^{i\theta_{+\bm{k}}}\,\cosh\eta_{+\bm{k}}\sinh\eta_{-\bm{k}}=-e^{i\theta_{\bm{k}}}\,\cosh\eta_{\bm{k}}\sinh\eta_{\bm{k}}\,,\label{E:fkgjhsd2}
\end{align}
where we have assumed~\cite{FDRSq,IntEnt} that the squeeze parameters $\eta_{\bm{k}}$, $\theta_{\bm{k}}$ and the rotation angle $\psi_{\bm{k}}$ depend only on the magnitude of $\bm{k}$. Eq.~\eqref{E:fkgjhsd1} gives the number density of the pair-created particles, while \eqref{E:fkgjhsd2} is a measure of coherence between the created particle~\cite{Par69}.

The Bogoliubov coefficients can be related to the fundamental solutions by~\cite{FDRSq,IntEnt}
\begin{align}
	\alpha_{\bm{k}}&=\frac{1}{2k}\Bigl[k\,d^{(1)}_{k}(\eta)+i\,d'^{(1)}_{k}(\eta)-i\,k^{2}\,d^{(2)}_{k}(\eta)+k\,d'^{(2)}_{k}(\eta)\Bigr]\,,\\
	\beta_{\bm{k}}&=\frac{1}{2k}\Bigl[k\,d^{(1)}_{k}(\eta)-i\,d'^{(1)}_{k}(\eta)-i\,k^{2}\,d^{(2)}_{k}(\eta)-k\,d'^{(2)}_{k}(\eta)\Bigr]\,.
\end{align}
Thus we readily find that for the mode function \eqref{E:fjbdkfs}, we have
\begin{align}
	\lvert\beta_{\bm{k}}(\eta)\rvert^{2}&=\frac{1}{8k^{8}\eta^{4}\eta_{0}^{4}}\Bigl\{1+2k^{4}\bigl(\eta^{4}+\eta_{0}^{4}\bigr)-\bigl[1-2k^{2}(\eta-\eta_{0})^{2}+4k^{4}\eta^{2}\eta_{0}^{2}\bigr]\cos2k(\eta-\eta_{0})\Bigr.\notag\\
	&\qquad\qquad\qquad\qquad\qquad-\Bigl.2k\bigl(\eta-\eta_{0}\bigr)\bigl(1+2k^{2}\eta\eta_{0}\bigr)\sin2k(\eta-\eta_{0})\Bigr\}>0\,.
\end{align}
It is oscillatory with frequency $2k$ when $\lvert2k\eta\rvert>1$, but the amplitude is proportional to roughly $k^{-4}$. On the other hand, if $\lvert2k\eta\rvert<1$ it gradually stops oscillating and transits to a monotonic increase as $\eta\to0^{-}$ like $k^{-4}\eta^{-4}$. That is, the number density of created particles of each mode oscillates with time when its physical wavelength is smaller than the horizon width, but when the physical wavelength becomes greater than the horizon, it grows monotonically.

Similar behavior is observed for the coherence between the created particles $C_{\bm{k}}=\alpha_{\bm{k}}^{\vphantom{*}}\beta^{*}_{\bm{k}}$. However two points are worth emphasizing. First, this quantity is not positive definite like the particle number density. And its phase will come to a constant $\pi$, as shown in Fig.~\ref{Fi:coherence}.
\begin{figure}
\centering
    \scalebox{0.5}{\includegraphics{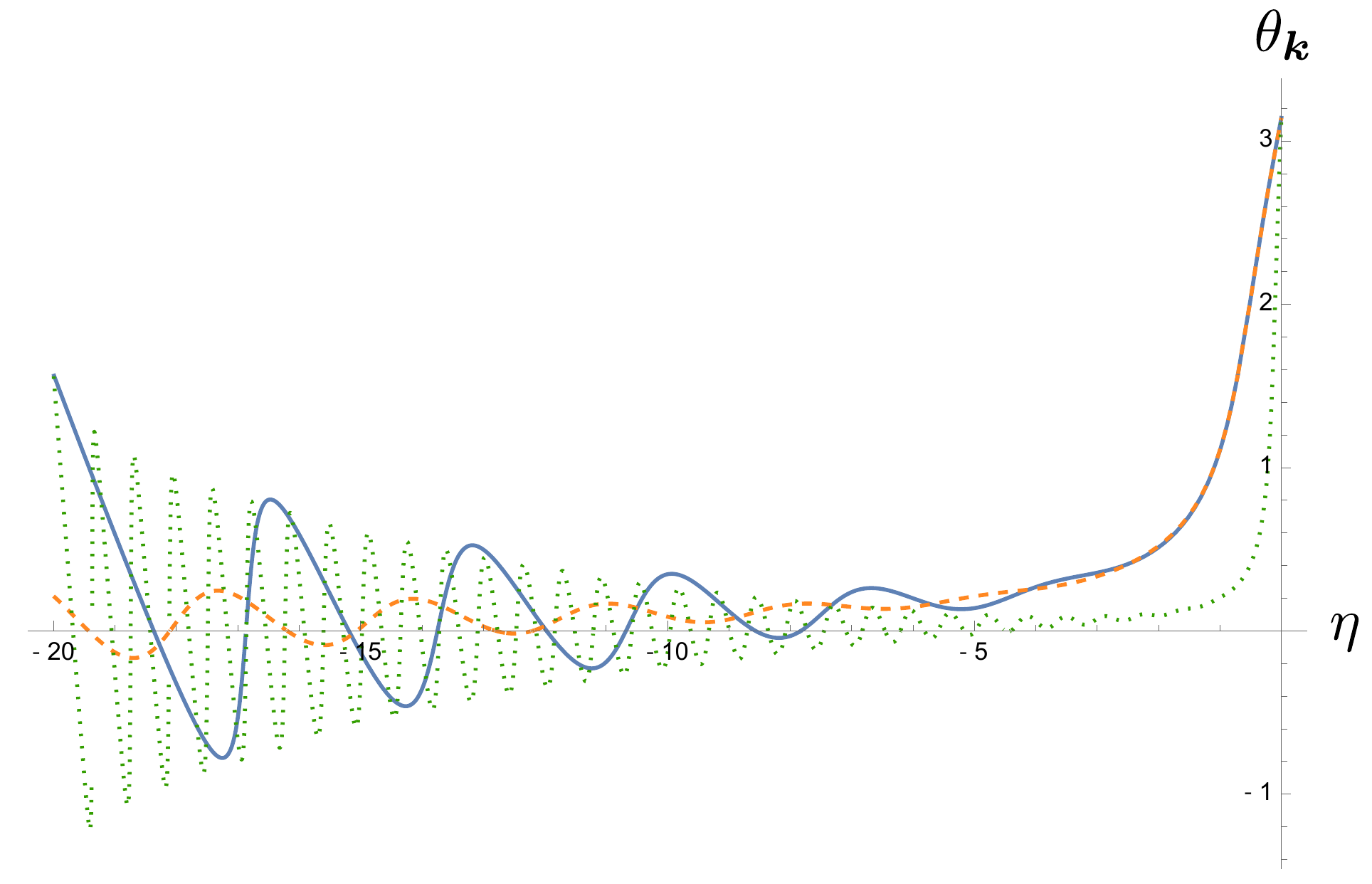}}
    \caption{The time variation of the phase $\theta_{\bm{k}}$ in coherence $C_{\bm{k}}=\lvert C_{\bm{k}}\rvert \,e^{i\theta_{\bm{k}}}$ for mode $\bm{k}$. The blue solid curve has the initial time is $\eta_{0}=-20$ and $k=1$, the orange dashed curve $\eta_{0}=-40$ and $k=1$, and the green dotted curve $\eta_{0}=-20$ and $k=5$.}\label{Fi:coherence}
\end{figure}
This can be seen from the expansion of $C_{\bm{k}}$ about $\eta=0^{-}$
\begin{equation}\label{E:fgbjehds}
	e^{i\theta_{\bm{k}}}\simeq-1-i\,2k\eta+2k^{2}\eta^{2}+\cdots\,.
\end{equation}
In Fig.~\ref{Fi:coherence}, it is interesting to note that for $\eta$ sufficiently close to $\eta_{0}$, the blue solid curve and the green dotted curve follow the same envelope. However, when $\lvert k\eta\rvert\ll1$, the blue curve more or less overlap with the red dashed curve, independent of the initial time, as has ben shown in \eqref{E:fgbjehds}. We also observe that the initial phase is $\pi/2$, seemingly contradictory to the fact that $\alpha_{\bm{k}}(\eta_{0})=1$ and $\beta_{\bm{k}}(\eta_{0})=0$. It may be resolved by their Taylor expansions around $\eta=\eta_{0}$ 
\begin{align}
	\alpha_{\bm{k}}(\eta)&=1+i\,\frac{1-k^{2}\eta_{0}^{2}}{k\eta_{0}^{2}}\,\bigl(\eta-\eta_{0}\bigr)+\cdots\,,\\
	\beta_{\bm{k}}(\eta)&=-i\,\frac{1}{k\eta_{0}^{2}}\,\bigl(\eta-\eta_{0}\bigr)+\cdots\,.
\end{align}
Thus the value of the phase angle in fact comes from the first-order contribution of $\beta_{\bm{k}}(\eta)$.

The covariance matrix elements are given by
\begin{align}
	b_{\bm{k}}&=\frac{1}{2}\,\langle\bigl\{\hat{\chi}_{\bm{k}}^{\vphantom{\dagger}}(\eta),\hat{\chi}^{\dagger}_{\bm{k}}(\eta)\bigr\}\rangle=u_{k}^{\vphantom{*}}(\eta)u_{k}^{*}(\eta)\notag\\
	&=\frac{1}{4k^{7}\eta^{2}\eta_{0}^{4}}\Bigl\{\bigl(1+k^{2}\eta^{2}\bigr)\bigl(1+2k^{4}\eta_{0}^{4}\bigr)-\bigl(1-k^{2}\eta^{2}+4k^{2}\eta\eta_{0}-2k^{2}\eta_{0}^{2}+2k^{4}\eta^{2}\eta_{0}^{2}\bigr)\,\cos2k(\eta-\eta_{0})\Bigr.\notag\\
	&\qquad\qquad\qquad\qquad\qquad-\Bigl.2k\bigl(\eta-\eta_{0}+k^{2}\eta^{2}\eta_{0}-2k^{2}\eta\eta_{0}^{2}\bigr)\,\sin2k(\eta-\eta_{0})\Bigr\}\,,\label{E:idxdf1}\\
	a_{\bm{k}}&=\frac{1}{2}\,\langle\bigl\{\hat{p}_{\bm{k}}^{\vphantom{\dagger}}(\eta),\hat{p}^{\dagger}_{\bm{k}}(\eta)\bigr\}\rangle=u'^{\vphantom{*}}_{k}(\eta)u'^{*}_{k}(\eta)\notag\\
	&=\frac{1}{4k^{7}\eta^{4}\eta_{0}^{4}}\Bigl\{\bigl(1-k^{2}\eta^{2}+k^{4}\eta^{4}\bigr)\bigl(1+2k^{4}\eta_{0}^{4}\bigr)\Bigr.\notag\\
	&\qquad-\bigl(1-3k^{2}\eta^{2}+k^{4}\eta^{4}+4k^{2}\eta\eta_{0}-4k^{4}\eta^{3}\eta_{0}-2k^{2}\eta_{0}^{2}+6k^{4}\eta^{2}\eta_{0}^{2}-2k^{6}\eta^{4}\eta_{0}^{2}\bigr)\,\cos2k(\eta-\eta_{0})\notag\\
	&\qquad\qquad-\Bigl.2k\bigl(\eta-\eta_{0}-k^{2}\eta^{3}+3k^{2}\eta^{2}\eta_{0}-k^{4}\eta^{4}\eta_{0}-2k^{2}\eta\eta_{0}^{2}+2k^{4}\eta^{3}\eta_{0}^{2}\bigr)\,\sin2k(\eta-\eta_{0})\Bigr\}\,,\label{E:idxdf2}\\
	c_{\bm{k}}&=\frac{1}{2}\,\langle\bigl\{\hat{\chi}_{\bm{k}}^{\vphantom{\dagger}}(\eta),\hat{p}^{\vphantom{\dagger}}_{\bm{k}}(\eta)\bigr\}\rangle=\frac{1}{2}\bigl[u^{\vphantom{*}}_{k}(\eta)u^{*}_{k}(\eta)\bigr]'\notag\\
	&=\frac{1}{4k^{7}\eta^{3}\eta_{0}^{4}}\Bigl\{-\bigl(1+2k^{4}\eta_{0}^{4}\bigr)+\bigl(1-2k^{2}\eta^{2}+4k^{2}\eta\eta_{0}-2k^{4}\eta^{3}\eta_{0}-2k^{2}\eta_{0}^{2}+4k^{4}\eta^{2}\eta_{0}^{2}\bigr)\,\cos2k(\eta-\eta_{0})\Bigr.\notag\\
	&\qquad\qquad\qquad+\Bigl.k\bigl(2\eta-2\eta_{0}-k^{2}\eta^{3}+4k^{2}\eta^{2}\eta_{0}-4k^{2}\eta\eta_{0}^{2}+2k^{4}\eta^{3}\eta_{0}^{2}\bigr)\,\sin2k(\eta-\eta_{0})\Bigr\}\,.\label{E:idxdf3}
\end{align}
As shown in Fig.~\ref{Fi:cov-2}, on the subhorizon scales, they oscillates with time, but once the physical wavelength of the mode crosses the Hubble horizon, they increase  monotonically and indefinitely. 

\begin{figure} 	
	\centering     	
	\scalebox{0.4}{\includegraphics{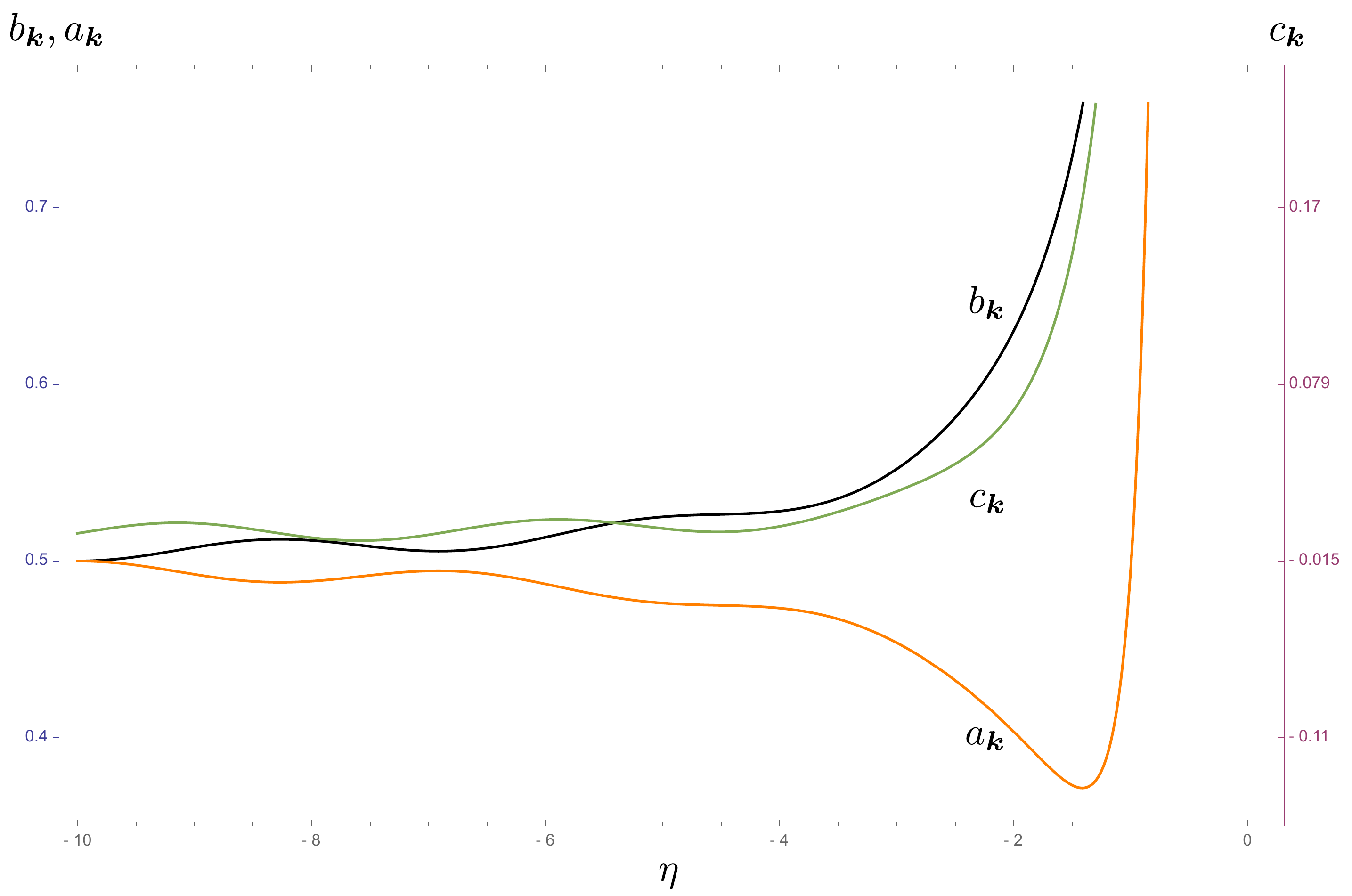}}\caption{{The time variation of the covariance matrix elements for mode $\bm{k}$. The blue solid curve denotes $b_{\bm{k}}=\frac{1}{2}\,\langle\bigl\{\hat{\chi}_{\bm{k}}^{\vphantom{\dagger}}(\eta),\hat{\chi}^{\dagger}_{\bm{k}}(\eta)\bigr\}\rangle$, the orange dashed curve $a_{\bm{k}}=\frac{1}{2}\,\langle\bigl\{\hat{p}_{\bm{k}}^{\vphantom{\dagger}}(\eta),\hat{p}^{\dagger}_{\bm{k}}(\eta)\bigr\}\rangle$, and the green dotted curve $c_{\bm{k}}=\frac{1}{2}\,\langle\bigl\{\hat{\chi}_{\bm{k}}^{\vphantom{\dagger}}(\eta),\hat{p}^{\vphantom{\dagger}}_{\bm{k}}(\eta)\bigr\}\rangle$. We choose the parameter $\eta_{0}=10$ and $k=1$. Hence their behaviors are qualitatively different, when $\lvert k\eta\rvert<1$, where the corresponding mode has a physical wavelength greater than the Hubble horizon.}}\label{Fi:cov-2}  
\end{figure}

Roughly there is a trend that $c_{\bm{k}}$ will lie between $b_{\bm{k}}$ and $a_{\bm{k}}$ in the super-horizon regime although it is not accurately portrayed in Fig.~\ref{Fi:cov-2} because different vertical scales are used. In addition, the element $c_{\bm{k}}(\eta)$ seems always positive for the superhorizon modes
\begin{equation}\label{E:dgeusdh1}
	c_{\bm{k}}(\eta)\simeq-\frac{1}{2k^{3}\eta^{3}}+\cdots\,,
\end{equation}
for $k\eta\ll1$, and the dominant contribution is independent of $\eta_{0}$. That is, for $\lvert k\eta_{0}\rvert\gg1$, we always have $c_{\bm{k}}(\eta)>0$ when $k\eta\ll1$. For comparison, in the superhorizon regime, we have
\begin{align}\label{E:dgeusdh2}
	b_{\bm{k}}(\eta)&\simeq\frac{1}{2k^{3}\eta^{2}}+\cdots\,,&a_{\bm{k}}(\eta)&\simeq\frac{1}{2k^{3}\eta^{4}}+\cdots\,.
\end{align}
The trend is then clearly seen. We emphasize again that similar to the inverted oscillator, if we use the exact, full expressions of the covariance matrix elements, we always have
\begin{equation}\label{E:dkgbsdfg}
	a_{\bm{k}}(\eta)b_{\bm{k}}(\eta)-c_{\bm{k}}^{2}(\eta)=\frac{1}{4}\,,
\end{equation}
a signature of the unitary evolution of the pure Gaussian state. If one uses the approximated expressions \eqref{E:dgeusdh1} and \eqref{E:dgeusdh2} for the superhorizon modes, one would find
\begin{equation}
	a_{\bm{k}}(\eta)b_{\bm{k}}(\eta)-c_{\bm{k}}^{2}(\eta)=0\,.
\end{equation}
If one uses this condition to argue for the classicalization of inflationary cosmological perturbations, one would miserably be doomed because it violates the Robertson-Schr\"odinger uncertainty relation
\begin{equation}
	a_{\bm{k}}(\eta)b_{\bm{k}}(\eta)-c_{\bm{k}}^{2}(\eta)\geq\frac{1}{4}\,.
\end{equation}

\subsection{Entanglement: an indelible  signifier of quantumness}
If each mode of the inflaton field perturbation starts in a vacuum state, then the evolution will lead it to a two-mode squeezed vacuum, which remains a pure state. To examine whether there exists quantum entanglement between states of modes $\pm\bm{k}$, an easily computable entanglement measure for such a bi-partite pure Gaussian state is their reduced von Neumann entropy~\cite{Addesso, Addesso1,LCH10}. This is the von Neumann entropy of the reduced density matrix, which is obtained after we coarse-grain one party of the bi-partite pure Gaussian state associated with modes $\pm\bm{k}$. As is shown in~\cite{LCH10,IntEnt}, the entanglement measure takes the form
\begin{align}\label{E:kjngskd}
	\mathcal{S}_{\bm{k}}&=\bigl(N_{\bm{k}}+1\bigr)\,\ln\bigl(N_{\bm{k}}+1\bigr)-N_{\bm{k}}\,\ln N_{\bm{k}}\,,
\end{align}
with the created particle number density being $N_{\bm{k}}=\lvert\beta_{\bm{k}}\rvert^{2}$. Following earlier discussions on $\lvert\beta_{\bm{k}}\rvert$, we learn that it would change monotonically once the $\pm\bm{k}$ mode crosses the horizon, meaning that the physical wavelength of the $\pm\bm{k}$ mode is greater than the horizon. Therefore the pair-created particles are unambiguously entangled even for the superhorizon modes. This is a very strong evidence that the inflaton field perturbation remains quantum-mechanical. Together with the arguments presented earlier, we can say that the inflaton field perturbation never drops its quantum nature, and decoherence does not occur during the unitary evolution of the perturbation. Similar consideration has been carried out by~\cite{MV16}, who investigate the quantum discord in the bi-partite state among the modes $\pm\bm{k}$. They showed that for a pure state, the quantum discord takes the same form as the reduced von Neumann entropy, discussed above. We concur with their conclusion that `\textit{... the CMB is placed in a state which is ``very quantum.'' This means that it is certainly impossible to reproduce all the correlation functions in a classical picture ...}'.

Finally, in this context we mention the take-home message of our recent companion paper~\cite{IntEnt}, that the entropy of the bipartite state for modes $\pm\bm{k}$ is zero throughout the unitary evolution because it continues to be a pure state and the created particles are entangled. If somehow we lose track of one partner of the pair in mode $+\bm{k}$ or $-\bm{k}$, we will end up having nonzero entropy given by \eqref{E:kjngskd}. Thus implies that entropy production in cosmological particle creation does not result from particle creation per se, but is a consequence of the loss of complete information of the field.

\section{Conclusion}

In this paper we adopt the Heisenberg picture to re-derive some earlier results pertaining to the classicalization issue  {of closed linear quantum systems} where the free particle~\cite{KP98} and the inverted oscillator~\cite{GuthPi85} models have been used  as  analogues to the quantum cosmological perturbations in inflationary universe. The advantage of the Heisenberg equations in treating the analog  models systems lies in its physical transparency and proximity to the classical equations of motion, so one can directly identify the dynamical features of the systems in question.

The classical or semi-classical limit of a closed quantum system is subtler than what one might think. The cut- and-dry rule of thumb of simply taking $\hbar\to0$ {alone may not be} the proper way to reach a classical limit {even though it is well-known that classical physics does not contain $\hbar$}.  Neither is the large $n$  limit alone, as prescribed by the correspondence principle: A highly excited system is often expected to show  classical behavior because the energy difference of the neighboring states are small compared with the mean energy of the system. It is often stated that the quantum, discrete nature becomes increasingly obscure and a classical description becomes viable. We use a harmonic oscillator as an example to show that taking the large $n$ (excitation number) limit {does not produce equivalent results as compared to a} vanishing $\hbar$. To refresh these basic points we have included a pedagogical derivation in Appendix~\ref{S:nxkje} to show the conditions how the Wigner function in a phase-space formulation of the quantum harmonic oscillator can, in the suitable semi-classical limit, describe the corresponding paths in phase space of a classical harmonic oscillator -- {namely, $n\to\infty$ and $\hbar\to0$ while keeping the total energy fixed} .

A closed system  should remain quantum mechanical throughout its  unitary evolution. For the un-confined linear quantum systems such as the free particle, inverted oscillator and cosmological perturbations in the inflationary universe, they have a common feature, that is,  the  dispersions of the canonical variables tend to grow unbounded over the unitary evolution, so that at late times the quadrature ellipse becomes extremely squeezed in one direction and stretched in another direction.  It is then often claimed or implicitly argued that in this limit the ellipse can reduce to a well-defined path in phase space, and the Wigner function will be proportional to a delta function that defines the path and the proportionality factor gives the classical probability of the system along the emergent phase-space path. Since in this argument $\hbar$ remains finite, the resulting Wigner function clearly violates the requirement that a proper Wigner function is bounded both from above and below. Further, the density matrix elements converted from this delta-function like Wigner function is unphysical: 1) It does not correspond to a pure state if the closed system starts in pure state; 2) It does not describe a mixed state either because the purity of the state is greater than unity, 3) the density matrix has negative eigenvalues, violating unitarity, leading to negative probability and ill-defined von Neumann entropy; And 4) the Robertson-Schr\"odinger uncertainty relation is not respected.

Furthermore,  regarding the evolution of the quantum operator, for unbounded motion, an approximation is often used where only the leading order contribution is kept. One would then show that the (canonical) operator at different times commute and even the equal-time commutation relation vanishes, and, voila,  classicality emerges. This is another gaffe in the folklore. As a matter of fact, if one considers the full contribution, even just keeping the subleading contribution, one can unambiguously demonstrate  that the aforementioned operator does not commute, and the commutation relation is preserved.

Therefore it is an oversimplification to regard the large squeezing limit as a classical limit because the quantum features of this closed system remain intact and discounting or dismantling them leads to ill-defined mathematical properties and unphysical consequences. {This is perspicuously  seen if we perform an unsqueezing via an unitary transformation. The quantum coherence of the system can be restored.} Finally, in cosmological perturbations, there is a more compelling argument for the preservation of quantumness. The particle pairs created during the inflationary universe are entangled. Entanglement is a uniquely quantum feature, and its existence is especially important if one wishes to trace back the quantum origin of cosmological perturbations using quantum information-theoretical tools.   \\

\noindent {\bf Acknowledgments} J.-T. Hsiang is supported by the Ministry of Science and Technology of Taiwan, R.O.C. under Grant No.~MOST 110-2811-M-008-522.

\clearpage
\appendix

\section{preservation of the uncertainty function under squeeze transformation}\label{S:ehhee}

Here we show that the squeeze transformation does not modify the bound in the generalized uncertainty relation for the free, linear quantum scalar field.  Since the free quantum scalar field in flat space can be viewed as a collection of quantum harmonic oscillators, we will use a harmonic oscillator to illustrate this point. Suppose that the oscillator is in an arbitrary normalized state $\lvert\psi\rangle$. The generalized uncertainty relation can be expressed as
\begin{align}
	\langle\hat{\chi}^{2}\rangle\langle\hat{p}^{2}\rangle-\frac{1}{4}\langle\bigl\{\hat{\chi},\hat{p}\bigr\}\rangle^{2}=C^{2}-AB
\end{align}
where the displacement and the conjugated momentum of the oscillator are respectively
\begin{align}
	\hat{\chi}&=\frac{1}{\sqrt{2m\omega}}\bigl(\hat{a}^{\dagger}+\hat{a}\bigr)\,,&\hat{p}&=i\frac{m\omega}{\sqrt{2}}\bigl(\hat{a}^{\dagger}-\hat{a}\bigr)\,,
\end{align}
such that
\begin{align*}
	\langle\hat{\chi}^{2}\rangle&=\frac{1}{2m\omega}\bigl(A+B+2C\bigr)\,,&\langle\hat{p}^{2}\rangle&=-\frac{m\omega}{2}\bigl(A+B-2C\bigr)\,,&\frac{1}{2}\langle\bigl\{\hat{\chi},\hat{p}\bigr\}\rangle&=\frac{i}{2}\bigl(B-A\bigr)\,,
\end{align*}
with $A=\langle\psi\rvert\hat{a}^{2}\lvert\psi\rangle$, $B=\langle\psi\rvert\hat{a}^{\dagger2}\lvert\psi\rangle$, and $C=\langle\psi\rvert\hat{a}^{\dagger}\hat{a}\lvert\psi\rangle+1/2$.

If we apply the squeeze operator $\hat{S}$ on the state $\lvert\psi\rangle$, then its associated actions on $\hat{a}$ can be given by
\begin{align}
	\hat{S}\hat{a}\hat{S}^{\dagger}=\mu\,\hat{a}+\nu\,\hat{a}^{\dagger}\,,
\end{align}
with $\mu^{2}-\lvert\nu\rvert^{2}=1$. We then have
\begin{align*}
	\hat{S}\hat{\chi}\hat{S}^{\dagger}&=\frac{1}{\sqrt{2m\omega}}\bigl[\bigl(\mu+\nu\bigr)\,\hat{a}^{\dagger}+\bigl(\mu+\nu^{*}\bigr)\,\hat{a}\bigr]\,,&\hat{S}\hat{p}\hat{S}^{\dagger}&=i\sqrt{\frac{m\omega}{2}}\bigl[\bigl(\mu-\nu\bigr)\,\hat{a}^{\dagger}-\bigl(\mu-\nu^{*}\bigr)\,\hat{a}\bigr]\,,
\end{align*}
such that
\begin{align}
	\langle\hat{S}\hat{\chi}^{2}\hat{S}^{\dagger}\rangle\langle\hat{S}\hat{p}^{2}\hat{S}^{\dagger}\rangle-\frac{1}{4}\langle\hat{S}\bigl\{\hat{\chi},\hat{p}\bigr\}\hat{S}^{\dagger}\rangle^{2}=\bigl(C^{2}-AB\bigr)\bigl(\mu^{2}-\lvert\nu\rvert^{2}\bigr){}^{2}&=C^{2}-AB\\
	&=\langle\hat{\chi}^{2}\rangle\langle\hat{p}^{2}\rangle-\frac{1}{4}\langle\bigl\{\hat{\chi},\hat{p}\bigr\}\rangle^{2}\,.\notag
\end{align}
It then shows that in general the squeezing does not modify the generalized uncertainty relation; it only distorts the quadratures in the relation.

\section{semiclassical limit of the harmonic oscillator}\label{S:nxkje}
{It is well-known that the motion of a classical harmonic oscillator traces out an ellipse in the phase space of its canonical variables $(x,p)$. Thus it is quite naturally to ask whether the Wigner function for a quantum harmonic oscillator in the phase space formulation of quantum mechanics can, in a suitable semi-classical limit, reveal  the same feature. That is, can one show that in the semi-classical limit, the Wigner function reduces to
\begin{equation}
	W_{n}(x,p)=\frac{1}{2\pi}\,\delta(\mathcal{H}-E_{n})\,,
\end{equation}
which is consistent with  classical expectation. The following discussion is largely based on the work~\cite{TZ00}, with notations adapted to this paper.  Suppose the oscillator has an energy $E_{n}=(n+1/2)\omega$, which is held fixed when  the semi-classical limit $\hbar\to0$ is reached. Thus the excitation number $n$ will grow accordingly.}

{It turns out more convenient to use the double Fourier transform, the characteristic function, of the Wigner function
\begin{align}
	C(\chi,\kappa)=\int\!dqdp\;e^{\frac{i}{\hbar}(\kappa q+p\chi)}\,W(q,p)\,.
\end{align}
In particular, the characteristic function $C(\chi,\kappa)$ of a pure state can be written as
\begin{align}
	C(\chi,\kappa)&=\int\!dqdp\;e^{\frac{i}{\hbar}(\kappa q+p\chi)}\frac{1}{2\pi\hbar}\int\!e^{-\frac{i}{\hbar}py}\psi(q+\frac{y}{2})\psi^{*}(q-\frac{y}{2})\notag\\
	&=\int\!dq\;e^{\frac{i}{\hbar}\,\kappa q}\,\psi(q+\frac{\chi}{2})\psi^{*}(q-\frac{\chi}{2})\,.
\end{align}
For the $n^{\text{th}}$ excited state $\psi_{n}(q)$ of the harmonic oscillator, 
\begin{align}\label{E:dngksjd}
	\psi_{n}(q)&=\frac{1}{\sqrt{2^{n}n!}}\biggl(\frac{\alpha^{2}}{\pi}\biggr)^{\frac{1}{4}}e^{-\frac{\alpha^{2}}{2}q^{2}}H_{n}(\alpha q)\,,&\alpha^{2}&=\frac{m\omega}{\hbar}\,,
\end{align}
we have
\begin{equation}
	C_{n}(\chi,\kappa)=\alpha\int\!dq\;e^{\frac{i}{\hbar}\,\kappa q}\,\frac{1}{\sqrt{\pi}2^{n}n!}\,e^{-\alpha^{2}(q^{2}+\frac{\chi^{2}}{4})}H_{n}[\alpha(q-\frac{\chi}{2})]H_{n}[\alpha(q+\frac{\chi}{2})]\,.
\end{equation}
Here $H_{n}(z)$ is the Hermite polynomial of order $n$. To evaluate this, we first form the generating function of $C_{n}(\chi,\kappa)$ by
\begin{align}
	\sum_{n=0}C_{n}(\chi,\kappa)\,t^{n}&=\alpha\int\!dq\;e^{\frac{i}{\hbar}\,\kappa q}\,\frac{t^{n}}{\sqrt{\pi}2^{n}n!}\,e^{-\alpha^{2}(q^{2}+\frac{\chi^{2}}{4})}H_{n}[\alpha(q-\frac{\chi}{2})]H_{n}[\alpha(q+\frac{\chi}{2})]\notag\\
	&=\frac{1}{1-t}\,\exp\Bigl[-\frac{1+t}{1-t}\frac{\kappa^{2}+\alpha^{4}\hbar^{2}\chi^{2}}{4\alpha^{2}\hbar^{2}}\Bigr]\,.\label{E:irhdfj}
\end{align}
with the help of 
\begin{align*}
	\sum_{n=0}^{\infty}\frac{s^{n}}{2^{n}n!}e^{-\frac{y^{2}}{2}}H_{n}(y)\,e^{-\frac{z^{2}}{2}}H_{n}(z)&=\frac{1}{\sqrt{1-s^{2}}}\,\exp\Bigl[-\frac{1+s^{2}}{2(1-s^{2})}y^{2}+\frac{2s}{1-s^{2}}\,yz-\frac{1+s^{2}}{2(1-s^{2})}z^{2}\Bigr]\,,
\end{align*}
for $\lvert r\rvert<1$. We then write \eqref{E:irhdfj} into
\begin{align}
	\sum_{n=0}C_{n}(\chi,\kappa)\,t^{n}=\frac{1}{1-t}\,\exp\Bigl[-\frac{\kappa^{2}+\alpha^{4}\hbar^{2}\chi^{2}}{4\alpha^{2}\hbar^{2}}\Bigr]\exp\Bigl[-\frac{2t}{1-t}\frac{\kappa^{2}+\alpha^{4}\hbar^{2}\chi^{2}}{4\alpha^{2}\hbar^{2}}\Bigr]\,,
\end{align}
with
\begin{align}
	\frac{\kappa^{2}+\alpha^{4}\hbar^{2}\chi^{2}}{2\alpha^{2}\hbar^{2}}&=\frac{K}{2\hbar\omega}\,,&&\text{and}&K&=\frac{\kappa^{2}}{2m}+\frac{m\omega^{2}}{2}\,\chi^{2}\,,
\end{align}
and compare it with the generating function of the Laguerre polynomials
\begin{equation}
	\sum_{n=0}L_{n}(z)t^{n}=\frac{1}{1-t}\,\exp\Bigl[-\frac{t}{1-t}\,z\Bigr]\,.
\end{equation}
We find
\begin{equation}
	C_{n}(\chi,\kappa)=e^{-\frac{K}{2\hbar\omega}}L_{n}(\frac{K}{\hbar\omega})\,.
\end{equation}
Now we explore an asymptotic expression of the generalized Laguerre polynomial $L^{(\alpha)}_{n}(z)$ for sufficiently large $n$,
\begin{align}
	L^{(\alpha)}_{n}(z)\simeq\frac{\Gamma(n+\alpha+1)}{n!}\biggl(\frac{4}{\nu z}\biggr)^{\frac{\alpha}{2}}\biggl[\frac{\varphi(t)}{\varphi'(t)}\biggr]^{\frac{1}{2}}\,\frac{e^{z/2}}{\sqrt{2t}}\,J_{\alpha}[\nu\varphi(t)]+\cdots\,,
\end{align}
valid for $z\lesssim \nu$, with
\begin{align}
	\nu&=4n+2\alpha+2\,,&t&=\frac{z}{\nu}\,,&\varphi(t)&=\frac{1}{2}\sqrt{t-t^{2}}+\frac{1}{2}\sin^{-1}\sqrt{t}\,,
\end{align}
and the Bessel function of first kind $J_{\alpha}(z)$. When $\alpha=0$, we have
\begin{equation}
	L_{n}(z)\simeq\biggl[\frac{\varphi(t)}{\varphi'(t)}\biggr]^{\frac{1}{2}}\,\frac{e^{z/2}}{\sqrt{2t}}\,J_{0}[\nu\varphi(t)]+\cdots\,,
\end{equation}
and $\nu=4n+2$. In the limit $t\ll1$, we find
\begin{align}
	\psi(t)&\simeq\sqrt{t}-\frac{t^{\frac{3}{2}}}{6}+\cdots\,,&\biggl[\frac{\varphi(t)}{\varphi'(t)}\biggr]^{\frac{1}{2}}&\simeq\sqrt{2t}+\frac{t^{\frac{3}{2}}}{3\sqrt{2}}+\cdots\,,
\end{align}
and thus
\begin{align}
	e^{-z/2}L_{n}(z)\simeq J_{0}(\sqrt{\nu z})\,,
\end{align}
which is valid for $z\lesssim n$, or with $z=K/(\hbar\omega)\lesssim n$
\begin{equation}\label{E:jthgdnb}
	\int\!dqdp\;e^{\frac{i}{\hbar}(\kappa q+p\chi)}\,W_{n}(q,p)=C_{n}(\chi,\kappa)\simeq J_{0}(2\sqrt{(n+1)\frac{K}{\hbar\omega}}\;)\,.
\end{equation}
Going in parallel, we check the double Fourier transformation of the delta function $\delta(\mathcal{H}-E_{n})$
\begin{align}\label{E:eindfs}
	\int\!dqdp\;e^{\frac{i}{\hbar}(\kappa q+p\chi)}\,\delta(\mathcal{H}-E_{n})=\int\!dqdp\;e^{\frac{i}{\hbar}(\kappa q+p\chi)}\,\delta(\frac{p^{2}}{2m}+\frac{m\omega^{2}}{2}q^{2}-E_{n})\,.
\end{align}
It is convenient to make the change of variables
\begin{align}
	Q&=\sqrt{\frac{m\omega^{2}}{2}}\;1\,,&P&=\frac{1}{\sqrt{2m}}\,p\,,&&\text{and}&x&=\frac{\sqrt{2m}}{\hbar}\,\chi\,,&k&=\frac{1}{\hbar}\sqrt{\frac{2}{m\omega^{2}}}\,\kappa\,,
\end{align}
such that \eqref{E:eindfs} becomes
\begin{align}
	\int\!dqdp\;e^{\frac{i}{\hbar}(\kappa q+p\chi)}\,\delta(\mathcal{H}-E_{n})&=\frac{2}{\omega}\int\!dQdP\;e^{i(Px+kQ)}\,\delta(Q^{2}+P^{2}-E_{n})\notag\\
	&=\frac{2}{\omega}\int_{0}^{2\pi}\!d\Theta\!\int_{0}^{\infty}dR\;R\,e^{i\lambda R\cos\Theta}\,\delta(R^{2}-E_{n})\notag\\
	&=\frac{1}{\omega}\int_{0}^{2\pi}\;e^{i\lambda \sqrt{E_{n}}\cos\Theta}\notag\\
	&=\frac{2\pi}{\omega}\,J_{0}(\lambda \sqrt{E_{n}})\,,
\end{align}
where
\begin{align}
	\lambda&=\sqrt{k^{2}+x^{2}}=\frac{2\sqrt{K}}{\hbar\omega}\,,&R&=\sqrt{Q^{2}+P^{2}}\,.
\end{align}
Thus we have
\begin{equation}\label{E:bvxsuafsf}
	\int\!dqdp\;e^{\frac{i}{\hbar}(\kappa q+p\chi)}\,\delta(\mathcal{H}-E_{n})=\frac{2\pi}{\omega}\,J_{0}(\,\frac{2\sqrt{K\,E_{n}}}{\hbar\omega}\;)\,.
\end{equation}
Following our previous arguments, we would like to fix $E_{n}=(n+\frac{1}{2})\hbar\omega$ while taking the limit $\hbar\to0$, so \eqref{E:bvxsuafsf} becomes 
\begin{equation}\label{E:bafsf}
	\int\!dqdp\;e^{\frac{i}{\hbar}(\kappa q+p\chi)}\,\delta(\mathcal{H}-E_{n})=\frac{2\pi}{\omega}\,J_{0}(2\sqrt{(n+1)\frac{K}{\hbar\omega}}\;)\,.
\end{equation}
Comparing \eqref{E:bafsf} with \eqref{E:jthgdnb}, we obtain that in the combined limits $n\gg1$, $\hbar\to0$ with a fixed $E_{n}=(n+1/2)\hbar\omega$
\begin{equation}\label{E:xzswe}
	W_{n}(q,p)=\frac{\omega}{2\pi}\,\delta(\frac{p^{2}}{2m}+\frac{m\omega^{2}}{2}q^{2}-E_{n})\,.
\end{equation}
Therefore we find that when {both limits are taken together}, the Wigner function reduces to a form that is consistent with the classical dynamics of the harmonics oscillator. We next examine the normalization condition
\begin{align}
	\int\!dqdp\;W_{n}(q,p)=\frac{\omega}{2\pi}\int\!dqdp\;\delta(\mathcal{H}-E_{n})&=\frac{\omega}{2\pi}\int\!dQdP\;\delta(Q^{2}+P^{2}-E_{n})\notag\\
	&=\frac{1}{\pi}\int_{0}^{2\pi}\!d\Theta\!\int_{0}^{\infty}dR\;R\,\delta(R^{2}-E_{n})\notag\\
	&=1\,,
\end{align}
and check that it is satisfied.}

\end{document}